\documentclass{emulateapj}
\usepackage{apjfonts,rotating}

\def\chandra{{\it Chandra\/}}

\def\genx{{\it Generation-X\/}}
\def\smartx{{\it SMART-X\/}}

\def\hst{{\it {\it HST}\/}}

\def\rosat{{\it ROSAT\/}}

\def\xmm{{\it XMM-Newton\/}}

\def\xray{\hbox{X-ray}}

\def\cdfs{\hbox{CDF-S}}
\def\cdfn{\hbox{CDF-N}}
\def\etal{{et\,al.}}

\def\ltsima{$\; \buildrel < \over \sim \;$}
\def\simlt{\lower.5ex\hbox{\ltsima}}
\def\gtsima{$\; \buildrel > \over \sim \;$}
\def\simgt{\lower.5ex\hbox{\gtsima}}
\def\kms{\ifmmode{~{\rm km~s^{-1}}}\else{~km s$^{-1}$}\fi}
\def\lsim{\lower0.3em\hbox{$\,\buildrel <\over\sim\,$}}
\def\gsim{\lower0.3em\hbox{$\,\buildrel >\over\sim\,$}}

\def\h2{H$_2$}
\def\flux{ergs~cm$^{-2}$~s$^{-1}$}
\def\lum{ergs~s$^{-1}$}
\def\intensity{ergs~cm$^{-2}$~s$^{-1}$~deg$^{-2}$}

\def\arcmin{\mbox{$^\prime$}}

\def\aap{A\&A}
\def\apj{ApJ}
\def\apjl{ApJL}
\def\apjs{ApJS}
\def\aj{AJ}
\def\mnras{MNRAS}
\def\araa{ARA\&A}

\def\sblim{$5.1 \times 10^{-18}$}
\def\hblim{$3.7 \times 10^{-17}$}
\def\fblim{$2.4 \times 10^{-17}$}
\def\uhblim{$4.6 \times 10^{-17}$}
\def\sbcxrb{$(8.15 \pm 0.58) \times 10^{-12}$}
\def\hbcxrb{$(1.73 \pm 0.23) \times 10^{-11}$}
\def\fbcxrb{$(2.54 \pm 0.24) \times 10^{-11}$}
\def\uhbcxrb{$(1.04 \pm 0.14) \times 10^{-11}$}
\def\sbden{27,800}
\def\hbden{10,500}
\def\uhbden{8,400}
\def\fbden{22,600}
\def\agnden{14,900}
\def\galden{12,700}

\begin{document}

\shortauthors{LEHMER ET AL.}
\shorttitle{X-ray Number Counts in the 4~Ms CDF-S}

%
\title{
The 4 Ms Chandra Deep Field-South Number Counts Apportioned by Source Class: Pervasive Active Galactic Nuclei and the Ascent of Normal Galaxies
}
%

\author{
B.~D.~Lehmer,\altaffilmark{1,2}
Y.~Q.~Xue,\altaffilmark{3,4}
W.~N.~Brandt,\altaffilmark{3,4}
D.~M.~Alexander,\altaffilmark{5}
F.E.~Bauer,\altaffilmark{6}
M.~Brusa,\altaffilmark{7}
A.~Comastri,\altaffilmark{8}
R.~Gilli,\altaffilmark{8}
A.~E.~Hornschemeier,\altaffilmark{2}
B.~Luo,\altaffilmark{9}
M.~Paolillo,\altaffilmark{10}
A.~Ptak,\altaffilmark{2}
O.~Shemmer,\altaffilmark{11}
D.~P.~Schneider,\altaffilmark{3,4}
P.~Tozzi,\altaffilmark{12}
\& C.~Vignali\altaffilmark{13}
}
\altaffiltext{1}{The Johns Hopkins University, Homewood Campus, Baltimore, MD 21218, USA}
\altaffiltext{2}{NASA Goddard Space Flight Centre, Code 662, Greenbelt, MD 20771, USA} 
\altaffiltext{3}{Department of Astronomy and Astrophysics, Pennsylvania State University, University Park, PA 16802, USA}
\altaffiltext{4}{Institute for Gravitation and the Cosmos, Pennsylvania State University, University Park, PA 16802, USA}
\altaffiltext{5}{Department of Physics, Durham University, Durham, DH1 3LE, UK}
\altaffiltext{6}{Pontificia Universidad Catolica de Chile, Departamento de Astronomia y Astrofisica, Casilla 306, Santiago 22, Chile}
\altaffiltext{7}{Max-Planck-Institut f\"ur Extraterrestrische Physik, Giessenbachstrasse, D-85748 Garching, Germany}
\altaffiltext{8}{INAF--Osservatorio Astronomico di Bologna, Via Ranzani 1, Bologna, Italy}
\altaffiltext{9}{Harvard-Smithsonian Center for Astrophysics, 60 Garden Street, Cambridge, MA 02138, USA}
\altaffiltext{10}{Dipartimento di Scienze Fisiche, Universita` Federico II di Napoli, Via Cinthia, 80126 Napoli, Italy}
\altaffiltext{11}{Department of Physics, University of North Texas, Denton, TX 76203}
\altaffiltext{12}{INAF--Osservatorio Astronomico di Trieste, Via Tiepolo 11, I-34131 Trieste, Italy}
\altaffiltext{13}{Universit\'a di Bologna, Via Ranzani 1, Bologna, Italy}

%
\begin{abstract}
%

We present \hbox{0.5--2~keV}, \hbox{2--8~keV}, \hbox{4--8~keV}, and
\hbox{0.5--8~keV} (hereafter, soft, hard, ultra-hard, and full bands,
respectively) cumulative and differential number counts ($\log N$--$\log S$)
measurements for the recently completed $\approx$4~Ms \chandra\ Deep
Field-South (\cdfs) survey, the deepest \xray\ survey to date.  We implement a
new Bayesian approach, which allows reliable calculation of number counts down
to flux limits that are factors of \hbox{$\approx$1.9--4.3} times fainter than
the previously deepest number-counts investigations.  In the soft band, the
most sensitive bandpass in our analysis, the $\approx$4~Ms \cdfs\ reaches a
maximum source density of $\approx$\sbden~deg$^{-2}$.  By virtue of the
exquisite \xray\ and multiwavelength data available in the \cdfs, we are able
to measure the number counts from a variety of source populations (active
galactic nuclei [AGNs], normal galaxies, and Galactic stars) and subpopulations
(as a function of redshift, AGN absorption, luminosity, and galaxy morphology),
and test models that describe their evolution.  We find that AGNs still
dominate the \xray\ number counts down to the faintest flux levels for all
bands and reach a limiting soft-band source density of
$\approx$\agnden~deg$^{-2}$, the highest reliable AGN source density measured
at any wavelength.  
We find that the normal-galaxy counts rise rapidly near the flux limits, and at
the limiting soft-band flux, reach source densities of
$\approx$\galden~deg$^{-2}$ and make up 46 $\pm$ 5\% of the total number
counts. 
The rapid rise of the galaxy counts toward faint fluxes, and significant
normal-galaxy contributions to the overall number counts, indicate that normal
galaxies will overtake AGNs just below the $\approx$4~Ms soft-band flux limit 
and will provide a numerically significant new \xray\ source population in
future surveys that reach below the $\approx$4~Ms sensitivity limit.  We show
that a future $\approx$10~Ms \cdfs\ would allow for a significant increase in
\xray\ detected sources, with many of the new sources being cosmologically
distant ($z \simgt 0.6$) normal galaxies.

%
\end{abstract}
%


%
\section{Introduction}
%

Deep extragalactic \xray\ surveys conducted with \chandra\ and \xmm\ have
resolved the vast majority of the \hbox{0.5--10~keV} cosmic \xray\ background (CXRB) and have
provided substantial new insight into the \xray\ point sources detected (see,
e.g., Brandt \& Hasinger 2005; Brandt \& Alexander~2010).  A fundamental
quantity used to characterize the extragalactic \xray\ source population is the
cumulative \xray\ number counts, which quantifies how the cumulative number of
\xray\ sources per unit area, $N$, increases with decreasing flux, $S$ (also
referred to as $\log N$--$\log S$; see, e.g., Brandt \etal\ 2001; Rosati \etal\
2002; Moretti \etal\ 2003; Bauer \etal\ 2004; Kim \etal\ 2007; Georgakakis
\etal\ 2008).  The number counts provide an observational constraint that must
be considered when constructing physical models that describe the \xray\
evolution of extragalactic sources in the Universe.  For example, successful
models including the supermassive black hole (SMBH) accretion history of the
Universe (e.g., Gilli \etal\ 2007; Treister \etal\ 2009) and the \xray\
evolution of normal galaxies (primarily driven by \xray\ binaries and hot gas;
see, e.g., Ranalli \etal\ 2003, 2005) must predict \xray\ source densities
consistent with the observed number counts.  

Due to its high angular resolution and low background, \chandra\ is the only
current \xray\ observatory capable of providing new constraining data in the
ultra-deep regime (below \hbox{0.5--8~keV} fluxes of a few~$\times
10^{-16}$~\flux), since the deepest \chandra\ surveys have now greatly
surpassed the \xmm\ confusion limit (see, e.g., Brandt \etal\ 2001; Giacconi
\etal\ 2001, 2002; Alexander \etal\ 2003; Luo \etal\ 2008; Xue \etal\ 2011).
As new \chandra\ surveys continue to probe the extragalactic \xray\ universe to
fainter depths, the number counts continuously rise as fainter \xray\
populations are revealed.  Naturally, it is at the faintest flux levels where
ultradeep \chandra\ surveys are probing new regions of discovery space and
classes of extragalactic sources that were poorly sampled at \xray\ energies in
the past (e.g., star-forming and passive galaxies, and obscured and
low-luminosity AGNs).  

Bauer \etal\ (2004; hereafter, B04) measured the number counts for \xray\ point
sources detected in the $\approx$2~Ms {\it Chandra} Deep Field-North (CDF-N)
and $\approx$1~Ms CDF-South (CDF-S), the deepest \chandra\ surveys at the time.
Using the available multiwavelength data, B04 were able to distinguish between
different \xray\ emitting populations (e.g., active galactic nuclei [AGNs],
normal galaxies, and Galactic stars), and measure their contributions to the
CXRB.  B04 found that, generally, AGNs make up exclusively the relatively
bright number counts.  However, at the faintest flux levels in the
\hbox{0.5--2~keV} bandpass, the most sensitive bandpass studied in the CDF
surveys, B04 showed that normal galaxies start to comprise an appreciable
fraction of the number counts ($\approx$25\% of sources with \hbox{$S_{\rm
0.5-2~keV} \approx$~[2--10]~$\times 10^{-17}$~\flux}). 

Recently, the deepest extragalactic \xray\ survey yet conducted, the \cdfs, has
reached a total exposure of $\approx$4~Ms (Xue \etal\ 2011; hereafter, X11).  X11
presented point-source catalogs and data products for the survey and provided
basic multiwavelength classifications for the 740 individually detected \xray\
sources in their main catalog.  This work has revealed that, at the faintest
flux levels, normal galaxies are playing an increasingly
important role in the new sources detected.  For example, at \hbox{0.5--2~keV}
fluxes below $\approx$$5 \times 10^{-17}$~\flux, $\approx$50\% of the \xray\
detected sources are classified as likely normal galaxies; below the current
detection limits, it is almost certain that the normal galaxy fraction
continues to increase (e.g., B04; Ranalli \etal\ 2005). 

In this paper, we present number counts for the new $\approx$4~Ms CDF-S,
focusing on the faint-flux regime (\hbox{0.5--2~keV} fluxes
$\simlt$$10^{-17}$~\flux).  In $\S$2, we describe a new Bayesian method for
computing number counts, which properly accounts for biases and measurement
uncertainties that are present in the important flux regime near the detection
limit.  In $\S$3, we utilize the available multiwavelength photometric data as
well as optical spectroscopic and photometric redshift catalogs to quantify,
with good reliability, the relative contributions to the number counts from
AGNs, normal galaxies, and Galactic stars.  We further break down the number
counts to quantify the contributions that subpopulations make to the AGN (with
redshift, intrinsic AGN column density, and \xray\ luminosity subpopulations)
and normal galaxy (with redshift and morphological subpopulations) number
counts.  We conclude $\S$3 by discussing how each population and
subpopulation contributes to the overall CXRB intensity.  In $\S$4, we use our
subpopulation number-counts estimates to compare with predictions from
phenomenological models that describe how the \xray\ emission from accreting SMBHs
and normal galaxies are expected to evolve with cosmic time.  In $\S$5, we use
these models to estimate directly number counts to flux levels below current
detection limits and highlight the prospects of deeper \xray\ surveys.  In
$\S$6, we summarize our findings.

Throughout this paper, we make use of the main point-source catalog and data
products provided by X11.  The Galactic column density for the \cdfs\ is $8.8
\times 10^{19}$~cm$^{-2}$ (Stark \etal\ 1992).  All of the \hbox{X-ray} fluxes
and luminosities quoted throughout this paper have been corrected for Galactic
absorption.  In the \xray\ band, we make use of four bandpasses:
\hbox{0.5--2~keV} (soft band [SB]), \hbox{2--8~keV} (hard band [HB]),
\hbox{4--8~keV} (ultra-hard band [UHB]), and \hbox{0.5--8~keV} (full band
[FB]).  Values of $H_0$ = 70~\hbox{km s$^{-1}$ Mpc$^{-1}$}, $\Omega_{\rm M}$ =
0.3, and $\Omega_{\Lambda}$ = 0.7 are adopted throughout this paper (e.g.,
Spergel \etal\ 2003).

%
\section{Methodology for Computing Number Counts}
%

Our primary goal is to evaluate the number counts across the entire
$\approx$465~arcmin$^2$ $\approx$4~Ms \cdfs\ to the faintest possible flux
levels.  In the faint-flux regime, computing number counts presents a
challenge, since these calculations must properly account for (1) the
non-negligible spatial variations in sensitivity across the \chandra\ image,
(2) incompleteness issues related to source detection algorithms (see $\S$2.1),
and (3) the Eddington bias.  The B04 investigation of the CDF number counts
implemented Monte Carlo simulations using an extrapolated faint-flux
number-count model to measure and correct for these biases.  Although this
method provides reasonable first-order corrections that account for the biases
near the flux limits, it does not optimize the input faint-flux model.  Such an
optimization would require several Monte Carlo runs with varying faint-flux
extrapolations, which is not feasible due to the computational reqirements of
this procedure.  Such a limitation will therefore introduce small systematic
errors in the number-count measurements near the flux limit if the wrong
faint-end number-count prior is chosen.

%
%
\begin{figure}
\figurenum{1}
\centerline{
\includegraphics[width=9.5cm]{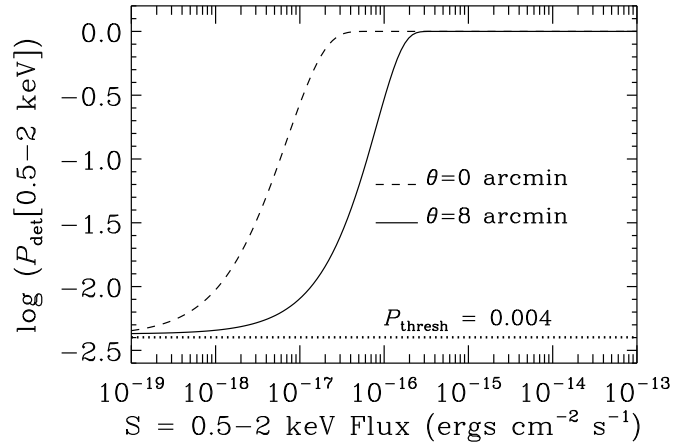}
}
\caption{
Detection probability as a function of intrinsic \hbox{0.5--2~keV} flux for
point sources located on axis ({\it dashed curve\/}) and $\approx$8\arcmin\ off
axis ({\it solid curve\/}).  These curves were computed considering a
counts-to-flux conversion factor appropriate for a power-law SED with $\Gamma =
1.4$ (see eqns.~2 and 3), the mean photon index of the CXRB.  Each curve
asymptotes to the defined detection probability threshold $P_{\rm thresh} =
0.004$ ({\it dotted line\/}) at low flux levels.  To construct the cumulative
solid angle as a function of flux limit for the entire $\approx$4~Ms CDF-S
image, we computed such curves at each location on the \chandra\ image
multiplied by the effective soild angle per pixel ($\approx$0.242~arcsec$^2$)
and added them (see Fig.~2).
}
\end{figure}

To mitigate this difficulty, we employ an approach similar to the
Poissonian-based methods described by Georgakakis \etal\ (2008; hereafter,
G08); however, our method has been adapted to account for the different
source-detection methods adopted for the $\approx$4~Ms \cdfs\ catalog by X11.
The X11 source-detection criteria make use of both {\ttfamily wavdetect}
(Freeman \etal\ 2002) for initial source selection and {\ttfamily ACIS Extract}
({\ttfamily AE}) to improve photometry and re-evaluate source-detection
significance (Broos \etal\ 2010).  {\ttfamily AE} makes use of binomial
statistics to evaluate source significance, and it properly accounts for PSF
variations between observations and uncertainties in local background
measurements (see below).  Our method for computing number counts, described in
detail below, uses a Bayesian approach with maximum-likelihood optimizations to
account for the Eddington bias and completeness limitations without requiring a
large number of time-consuming simulations. 

\subsection{Source Recovery Functions and Flux Probability Distributions}

Number-counts computations at the flux limits depend sensitively on the solid
angle of the survey over which a source of a given flux could be detected.  As
described in $\S$3 of X11, to be included in the main catalog, an \xray\ source
must (1) be detected by {\ttfamily wavdetect} at a false-positive probability
threshold of $10^{-5}$ and (2) contain $s$ counts (derived from {\ttfamily AE};
see column~8--16 of Table~3 in X11) within an aperture representing
$\approx$90\% of the point-source encircled-energy fraction (EEF)\footnote{As
noted by X11, in special cases where sources had nearby neighbors, the
source-detection probability was measured using an aperture smaller than the
$\approx$90\% EEF; however, only $\approx$2--3\% of all sources had moderate
source crowding, where detection probabilities were measured using apertures
that encompass $\simlt$50\% EEF.  We therefore expect that this will have a
negligible effect on our number-counts measurements.} that satisfies the
following binomial probability criterion:
\begin{equation}
P(x \ge s) = \displaystyle\sum\limits_{x=s}^{n} \frac{n!}{x! (n-x)!} p^x (1 -
p)^{n-x} \le P_{\rm thresh},
\end{equation}
where $n \equiv s + b_{\rm ext}$ and $p \equiv 1/(1+b_{\rm ext}/b_{\rm src})$.
Here $b_{\rm ext}$ is the total number of background counts extracted from a
large region outside of the point source (while masking out regions from other
\xray\ detected sources) that was used to obtain an estimate of the local
background count rate.  The quantity $b_{\rm src}$ is the estimated number of
background counts within the source extraction region, which was measured by
rescaling $b_{\rm ext}$ to the area of the source aperture (typically, $b_{\rm
ext}/b_{\rm src} \approx 16$; see $\S$4.1 of X11 for details).  In equation~1,
$P_{\rm thresh} = 0.004$, the value adopted by X11.  This choice of $P_{\rm
thresh}$ was empirically selected both to optimize the number of sources
detected and to ensure that nearly all detected sources are reliable (see
$\S$4.1 of X11).  This multi-stage procedure for identifying a highly reliable
list of source candidates will {\itshape not} result in a complete selection of
{\it all} real sources in the image having binomial probabilities $\le P_{\rm
thresh}$, since the initial {\ttfamily wavdetect} selection has more complex
source-detection criteria (see Freeman \etal\ 2002 for details) than the simple
criterion given in equation~1 (see $\S$2.2 and the Appendix for discussion on
correcting for these issues when computing number counts).  

%
%
\begin{figure}
\figurenum{2}
\centerline{
\includegraphics[width=9.5cm]{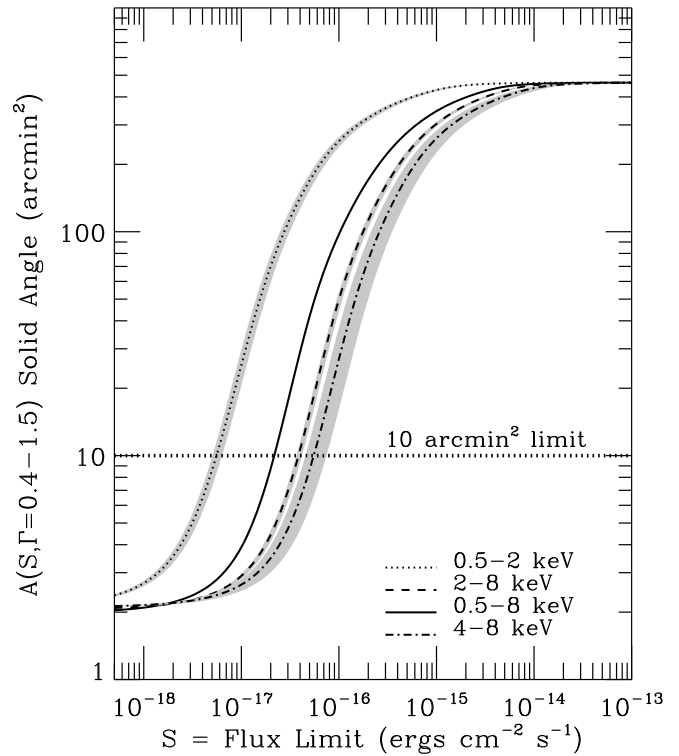}
}
\caption{
Effective solid angle $A(S)$ versus intrinsic flux limit $S$ for the SB ({\it dotted curve\/}),
HB ({\it short-dashed curve\/}), UHB ({\it dashed-dotted curve\/}), and FB
({\it solid curve\/}).  These curves have been computed for a power-law SED
with $\Gamma = 1.0$, the median photon index for sources detected in both the
SB and HB. The shaded regions give the effective solid angle curves appropriate for
the interquartile ($\approx$25--75\%) range $\Gamma =$~0.4--1.5.  These curves
were computed following the probabilistic methods discussed in $\S$2.1 and
account for uncertainties in measured flux conversions at the lowest flux
levels where the number of detected counts are small.  The horizontal dotted line
represents the $\approx$10~arcmin$^2$ limit, above which we can confidently
compute number counts (see $\S$2.1 for details).  These curves are
qualitatively different in nature from those produced by X11, which utilize
single-valued count-rate to flux conversions.
}
\end{figure}

Using a probabilistic approach, we compute the solid angle of the sky within
which a source with intrinsic flux $S$ would be detected if present.  First,
using eqn.~(1), we define the quantity $L$, which is the number of counts
required for a detection, which satisfies the relation $P(x \ge L) = P_{\rm
thresh} = 0.004$.  In order to compute $L$ at each location in the \cdfs\
image, it is necessary to estimate local values of $b_{\rm src}$ and $b_{\rm
ext}$ that are similar to those that would be estimated following the X11
approach used for the \xray\ detected sources.  Unfortunately, the X11
procedure for extracting these values at every location in the image is
computationally prohibitive, since source photometry is performed on an
observation-by-observation basis (i.e., in up to 54 observations in total) and
makes use of polygonal apertures that approximate local PSFs (see X11 for
further details).  To overcome this issue, we estimated $b_{\rm src}$ by
extracting background counts from the merged \cdfs\ background maps (see
$\S$7.1 of X11) using circular apertures with sizes that encompass the 90\% EEF
for a point source.  This approach was tested by comparing our $b_{\rm src}$
values measured in the regions of the X11 sources with those of X11.  We find
good agreement between values and a 1$\sigma$ scatter at the $\approx$18--25\%
level.  To estimate appropriate values of $b_{\rm ext}$ at each location
(pixel) on the \cdfs\ image, we measured the local off-axis angle $\theta_p$;
using the X11 point-source catalog, we adopted the maximum value of $b_{\rm
ext}$ for sources with off-axis angles $\theta = \theta_p \pm 0\farcm25$.
Given values of $b_{\rm src}$ and $b_{\rm ext}$, we numerically solved the
relation $P(x \ge L) = 0.004$ to obtain $L$ at each image location.  In this
manner, we constructed a spatial sensitivity map consisting of $L$ values.

In principle, $L$ could be combined with the exposure time to estimate a
count-rate limit, which can in turn be converted to a single flux limit using a
count-rate to flux conversion (as done in the X11 catalogs); however, this
single flux-limit approach does not directly incorporate the probabilistic
nature of source detection that is important in the low-count regime.  It is
therefore more informative to compute the probability that a source with
intrinsic flux, $S$, would be detected given that $L$ counts are required.
Such a source is expected to contribute the following number of counts to the
source detection cell:
\begin{equation}
T = t_{\rm exp} C \eta S  + b_{\rm src}
\end{equation}
where $t_{\rm exp}$, $C$, and $\eta$ are the effective exposure time, the
conversion from flux to count-rate, and the encircled-energy fraction,
respectively.  Values of $t_{\rm exp}$ are taken directly from the exposure
maps from X11 (see their $\S$3.1).  These exposure maps include the effects of
vignetting, gaps between CCDs, bad-column filtering, bad-pixel filtering, and
spatial and time-dependent degradation in quantum efficiency due to
contamination on the ACIS optical-blocking filters.  Values of $C$ will depend
on the spectral energy distribution (SED) shape.  In this work we use power-law
SEDs to characterize the observed spectra; these SEDs can be described using
the photon index $\Gamma$.  For each source, $\Gamma$ was derived using the
HB-to-SB count-rate ratio (corrected for differential vignetting and exposure
times) as a proxy for spectral slope.  The HB-to-SB count-rate ratio was
calibrated against $\Gamma$ using the AE-automated XSPEC-fitting procedure for
relatively bright X-ray sources (with full-band counts greater than 200; this
ensures reliable XSPEC-fitting results).  This approach takes into account the
multi-epoch Chandra calibration information (see X11 do further details).

Therefore, at each location on the image, the probability of source detection
as a function of flux $S$ can be computed following
\begin{equation}
P_{\rm det} =  \displaystyle\sum\limits_{x=L}^{L + b_{\rm ext} } \frac{(L +
b_{\rm ext})!}{x! (L + b_{\rm ext} - x)!} \left(\frac{T}{T + b_{\rm ext}}
\right)^x \left(1 - \frac{T}{T + b_{\rm ext}} \right)^{L+b_{\rm ext} - x}
\end{equation}
where $b_{\rm ext}$ is the extracted background counts that were used to
calculate the local background in eqn.~(1).  Figure~1 shows $P_{\rm det}$ as a
function of \hbox{0.5--2~keV} flux $S_{\rm 0.5-2~keV}$ for randomly selected
positions at $\theta \approx 0$\arcmin\ and 8\arcmin\ offsets from the average
aimpoint of the 4~Ms CDF-S (assuming $\eta = 0.9$ and a counts-to-flux
conversion factor appropriate for a power-law SED with $\Gamma = 1.4$).  We
note that as the flux drops to zero, $P_{\rm det}$ asymptotically approaches
our detection threshold $P_{\rm thresh} = 0.004$.  This shows that, even when
no source is present, there is still a finite (yet small, i.e., $P = 0.004$)
probability that a positive fluctuation may exceed our adopted detection
threshold; however, as shown by X11 through multiwavelength counterpart
matching, the initial {\ttfamily wavdetect} source selection ensures that very
few false sources are present in the X11 catalog.

%
%
\begin{figure}
\figurenum{3}
\centerline{
\includegraphics[width=9.5cm]{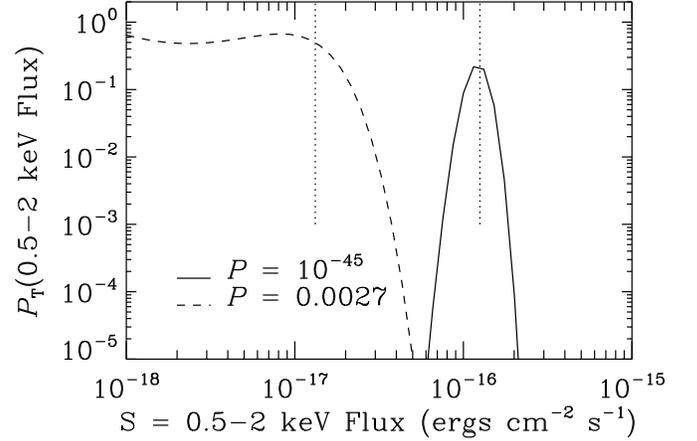}
}
\caption{
0.5--2~keV flux probability distributions for two AGNs in the X11 catalog with
no-source detection probabilities (provided by {\ttfamily AE}) of
$\approx$$10^{-45}$ ({\it solid curve\/}) and $\approx$$0.003$ ({\it dashed
curve\/}).  The latter source is roughly at the boundary of our adopted
detection threshold $P_{\rm thresh} = 0.004$, and therefore illustrates the
extent to which the flux conversion for sources in the low-count regime with
non-negligible Eddington biases can be affected.  The vertical dotted lines
indicate the most probable single-value flux given in the X11 catalog.
}
\end{figure}

We consider that each \chandra\ pixel represents a small local solid angle of
size $d\Omega \approx 0.242$~arcsec$^2$ (i.e.,
$\approx$0\farcs492~$\times$~0\farcs492) with a flux and SED (i.e., $\Gamma$)
dependent detection probability distribution characterized by $P_{\rm det}$.
Under this consideration, the total surveyed solid angle over which sources
with flux $S$ and photon index $\Gamma$ could be detected when present is
therefore $A(S,\Gamma) = \sum_i P_{{\rm det},i} \; d\Omega$ (hereafter, the
effective solid angle), where the summation is over all possible detection
cells (i.e., all pixels).  In Figure~2, we show $A(S, \Gamma = 1.0)$ versus $S$
for the $\approx$4~Ms \cdfs\ in the four bandpasses.  The shaded regions show
$A(S,\Gamma)$ in the range of $\Gamma =$~0.4--1.5 (median value $\Gamma =
1.0$), which represents the interquartile (i.e., 25--75\%) range for the 332
X11 sources with estimates of $\Gamma$ that were not based on limits.  We note
that generally $A(S,\Gamma)$ increases with increasing $\Gamma$.  We find that
the $A(S,\Gamma)$ curves asymptote to a value of $\approx$2~arcmin$^2$
approaching $S = 0$, suggesting that sources with extremely low fluxes (down to
zero) could in principle produce fluctuations exceeding the probability
threshold defined in equation~1.  However, as discussed above, such sources
would most efficiently be removed using our initial {\ttfamily wavdetect}
screening.  Therefore, we cannot use information at such flux levels to
determine number counts reliably.  We therefore choose to restrict our number
counts computations to flux levels where $>$10~arcmin$^2$ solid angle is
accessible in our survey for the case of $\Gamma = 1.4$ (the mean SED of the
\xray\ background; e.g., Moretti \etal\ 2009).  Our adopted
$\approx$10~arcmin$^2$ solid-angle limit additionally constrains our flux
limits.  The resulting limits are \sblim, \hblim, \uhblim, \fblim~\flux\ for
the SB, HB, UHB, and FB, respectively; these are factors of
\hbox{$\approx$1.9--4.3} times fainter than those of B04 and G08, the
previously deepest number-counts studies.  For the full range of \xray\
spectral slopes (i.e., $\Gamma$ values) in the X11 sample,
$\approx$5--20~arcmin$^2$ of solid angle is available for number counts
computations at the flux limits (above the asymptotic regime).  We note that
the flux limits derived here are fainter than those presented in X11, which
were $\approx$$9.1 \times 10^{-18}$ and $\approx$$5.5 \times 10^{-17}$~\flux\
in the SB and HB, respectively (see also the area curves in our Fig.~2 compared
with Fig~23 in X11).  This is due to the fact that X11 considered only a single
count-rate to flux conversion factor and did not use the probabilistic approach
adopted here.

To account for the fact that, for each \xray\ detected source in our main
catalog, we are only able to measure reliably the total observed counts $s$
(see above), the local background $b_{\rm ext}$, and that the intrinsic flux
$S$ may be subject to large uncertainty (particularly in the low-count regime),
we consider the conversion from counts to flux for each source to be
probabilistic.  For each \xray\ detected source, we computed the flux
probability distribution following
\begin{equation}
P_{\rm T} = \frac{(s + b_{\rm ext})!}{s! b_{\rm ext}!} \left(\frac{T}{T+b_{\rm
ext}} \right)^s \left( 1 - \frac{T}{T+b_{\rm ext}} \right)^{b_{\rm ext}}
dN/dS\big|_{\rm model}
\end{equation}
where the term $dN/dS\big|_{\rm model}$ is a Bayesian prior, based on the
differential number counts, which accounts for the Eddington bias near the
sensitivity limit.  
As noted in B04, the slope of the number counts of AGNs, normal galaxies, and
Galactic stars will differ at the flux limit of the \cdfs.  Therefore,
$dN/dS\big|_{\rm model}$ for a given source will depend on which source
population it belongs to.  Previous studies of \xray\ number counts (e.g.,
Rosati \etal\ 2002; B04; Kim \etal\ 2007; G08) have shown that power laws
provide good fits to the overall shapes of the $\log N$--$\log S$.  To first
order, we use priors based on the following power-law parameterizations:
\[\frac{dN}{dS}^{\rm AGN} = \left\{ 
\begin{array}{c c}
K^{\rm AGN} (S/S_{\rm ref})^{-\beta_1^{\rm AGN}}  & (S \le f^{\rm AGN}_{\rm b})
\\
K^{\rm AGN} (f_{\rm b}/S_{\rm ref})^{\beta_2^{\rm AGN} - \beta_1^{\rm AGN}}(S/S_{\rm
ref})^{-\beta_2^{\rm AGN}} &  (S > f^{\rm AGN}_{\rm b}) \\ \end{array} \right.
\]

\begin{eqnarray}
\frac{dN}{dS}^{\rm gal} & = & K^{\rm gal} (S/S_{\rm
ref})^{-\beta^{\rm gal}} \nonumber \\
\frac{dN}{dS}^{\rm star} & = & K^{\rm star} (S/S_{\rm
ref})^{-\beta^{\rm star}},
\end{eqnarray}
where $dN/dS({\rm AGN})$, $dN/dS({\rm gal})$, and $dN/dS({\rm star})$ are
differential number-count parameterizations to be applied to AGNs, normal
galaxies, and Galactic stars, respectively, $f_{\rm b}^{\rm AGN}$ is the flux
related to the break in the double power-law used to describe the AGN number
counts, and $S_{\rm ref} \equiv 10^{-14}$~\flux.  As we show in $\S$3.1 below,
we characterize each \xray\ source using the \xray\ and multiwavelength data
and provide best estimates of the parameters in equation~5 (i.e., $K$, $\beta$,
and $f_b$ values) for the AGNs, normal galaxies, and Galactic stars.  For each
source, we computed $P_{\rm T}$ down to the flux limits defined above and
normalize equation~4 using $\int P_T \; dS = 1$.  In Figure~3, we show examples
of $P_{\rm T}$ as a function of \hbox{0.5--2~keV} flux for two AGNs in the main
\chandra\ catalog having {\ttfamily AE} probability $P \approx 10^{-45}$ ({\it
solid curve\/}) and $P \approx 0.003$ ({\it dashed curve}; near the detection
threshold).  For this computation, we utilized values of $\beta_1^{\rm AGN} =
1.49$, $\beta_2^{\rm AGN} = 2.49$, and $f_{\rm b}^{\rm AGN} = 5.6 \times
10^{-15}$~\flux\ (the normalization is arbitrary); as we will show in $\S$3.1,
these values represent the best-fit parameterizations for AGN SB number counts.

%
%
\begin{figure} \figurenum{4} \centerline{
\includegraphics[width=8cm]{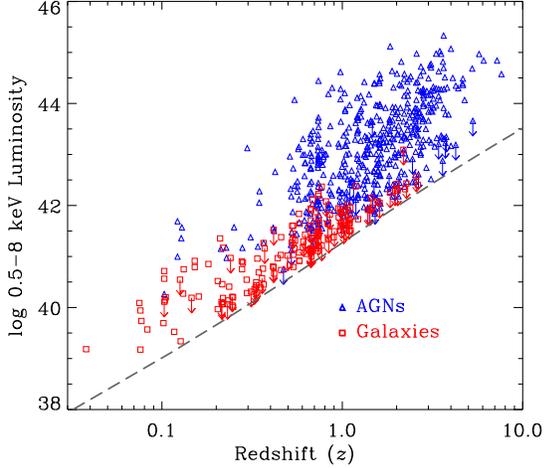} } \caption{
Observed-frame 0.5--8~keV luminosity versus redshift for AGNs ({\it blue
triangles\/}) and normal galaxies ({\it red squares\/}).  Luminosities were
calculated using the best redshift estimates and 0.5--8~keV fluxes provided by
X11 and our adopted cosmology.  The dashed curve indicates the observed-frame
luminosity corresponding to the flux limit at the center of the $\approx$4~Ms
\cdfs.  The normal-galaxy popluation broadly covers the redshift range \hbox{$z
\approx 0.03$--2.6} and AGNs cover the redshift range of \hbox{$z
\approx$~0.1--8}.  These redshifts are based on both spectroscopic and
photometric redshifts, and all sources at $z > 4.76$ have only photometric
redshift estimates.  We adopt the most probable photometric redshifts;
however, some of the $z> 4.76$ photometric redshifts are consistent with being
at lower redshifts (see Luo \etal\ 2010 for details).
} \end{figure}

%
%
\begin{figure*}
\figurenum{5}
\centerline{
\includegraphics[width=16cm]{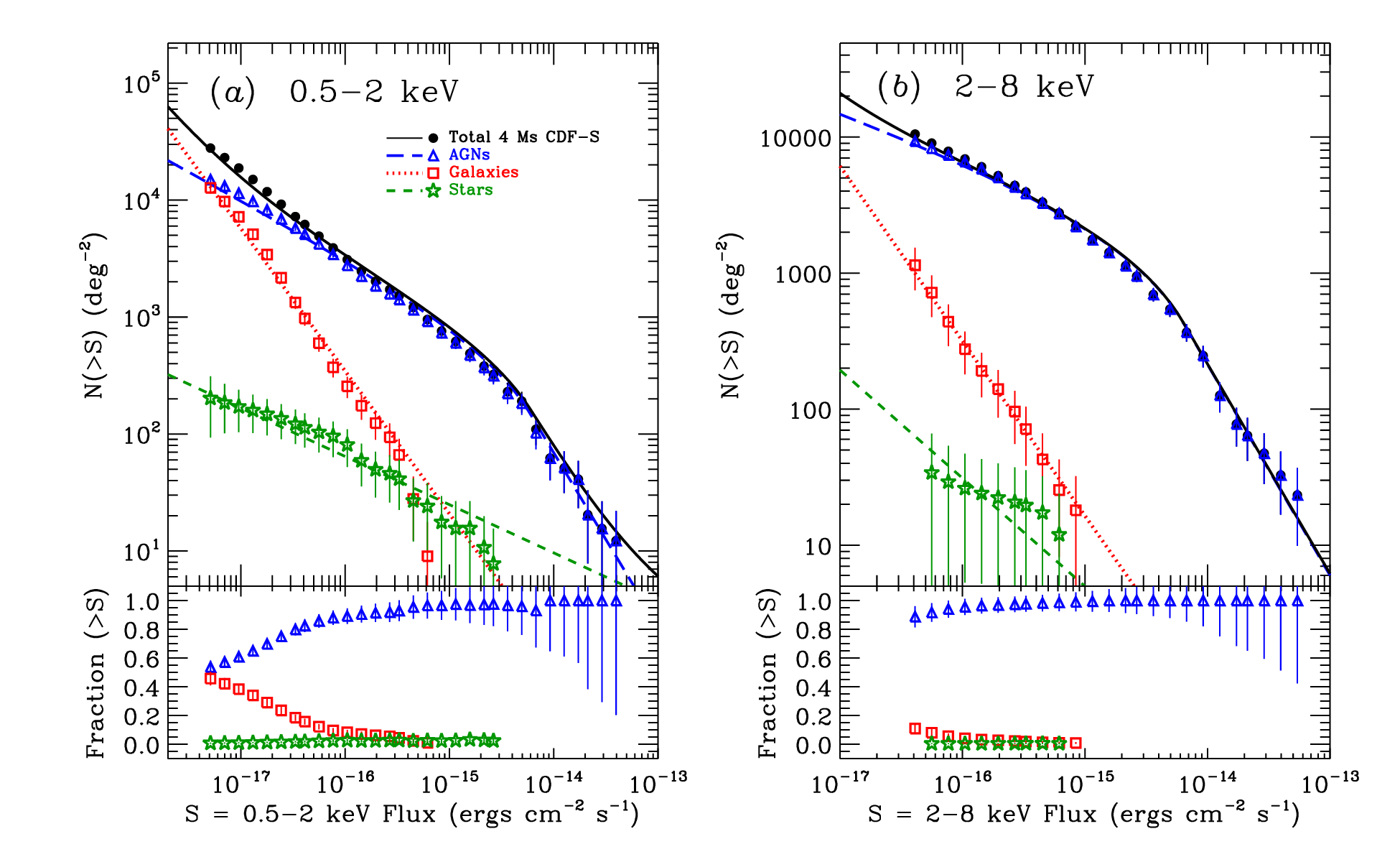}
}
\centerline{
\includegraphics[width=16cm]{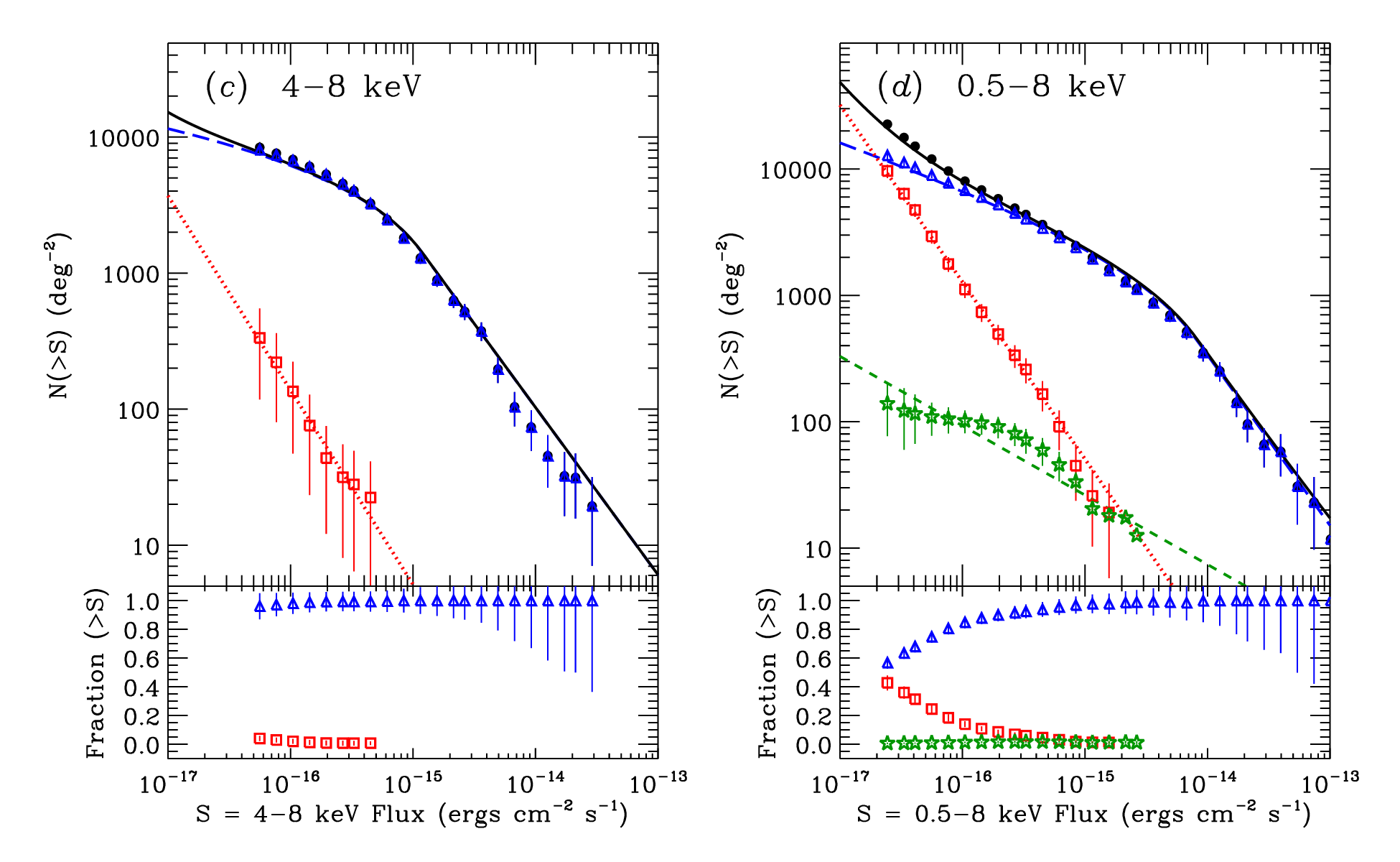}
}
\caption{
{\it Top panels of a--d}: Cumulative number counts for the SB ($a$), HB ($b$),
UHB ($c$), and FB ($d$) broken down into AGNs ({\it open blue triangles\/}), normal
galaxies ({\it open red squares\/}), and Galactic stars ({\it open green stars\/}).
Total number counts have been shown as filled circles.  In each plot, the
best-fit $dN/dS$ parameterizations based on equation~5 have been shown as blue
long-dashed, red dotted, and green short-dashed curves for AGNs, normal
galaxies, and Galactic stars, respectively.  The total number counts model,
based on summing the three contributing components, has been shown as a black
curve.
{\it Bottom Panels of a--d}: Fractional contributions from AGNs, normal
galaxies, and Galactic stars to the total number counts.  For the majority of
the flux ranges, AGNs dominate the number counts; however, normal galaxies
provide significant contributions near the flux limits of the SB and FB.
[{\it A machine-readable table of the cumulative number counts data for the $\approx$4~Ms CDF-S is
provided in the electronic edition.}]
}
\end{figure*}

%
%
\begin{figure*}
\figurenum{6}
\centerline{
\includegraphics[width=16cm]{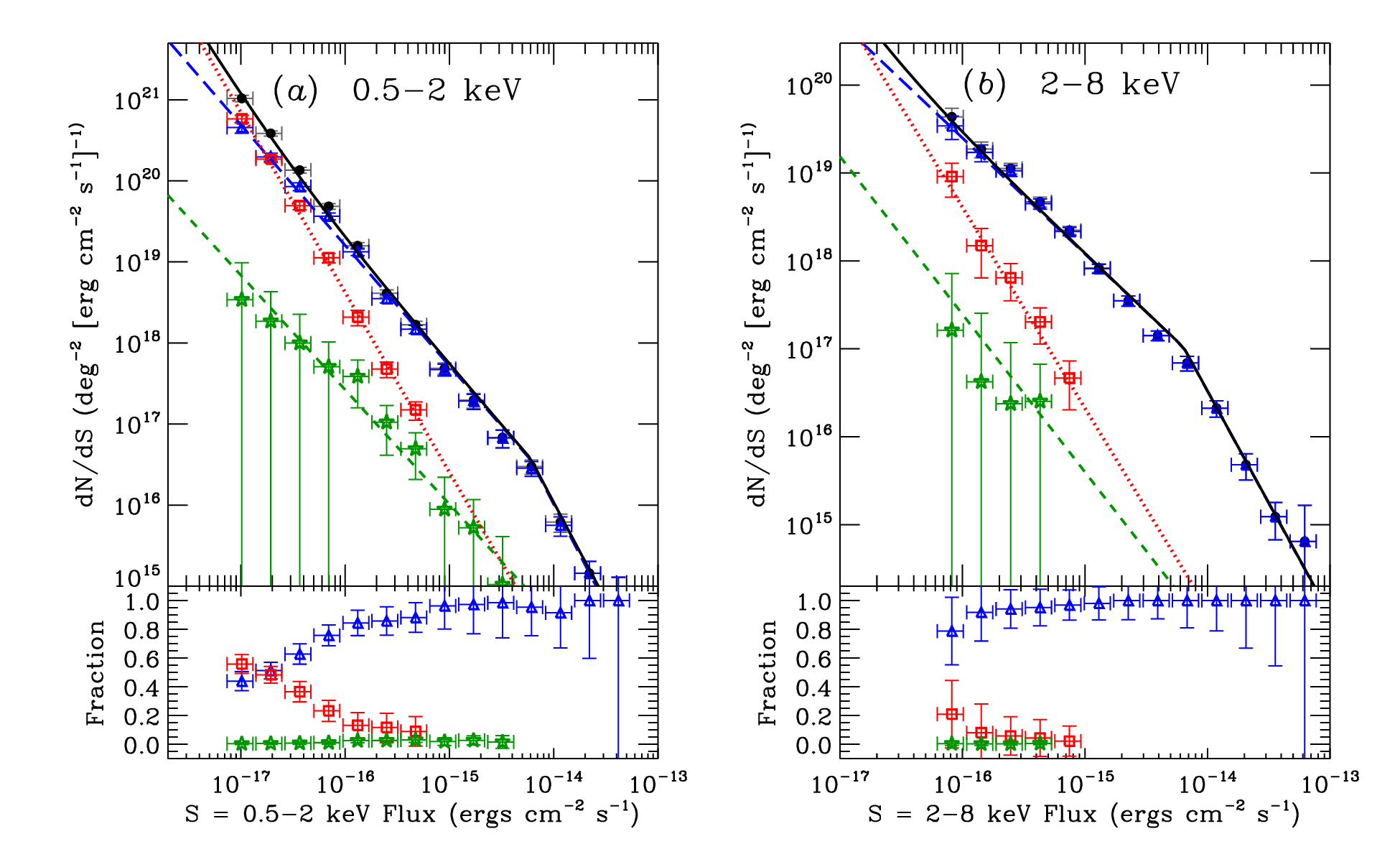}
}
\centerline{
\includegraphics[width=16cm]{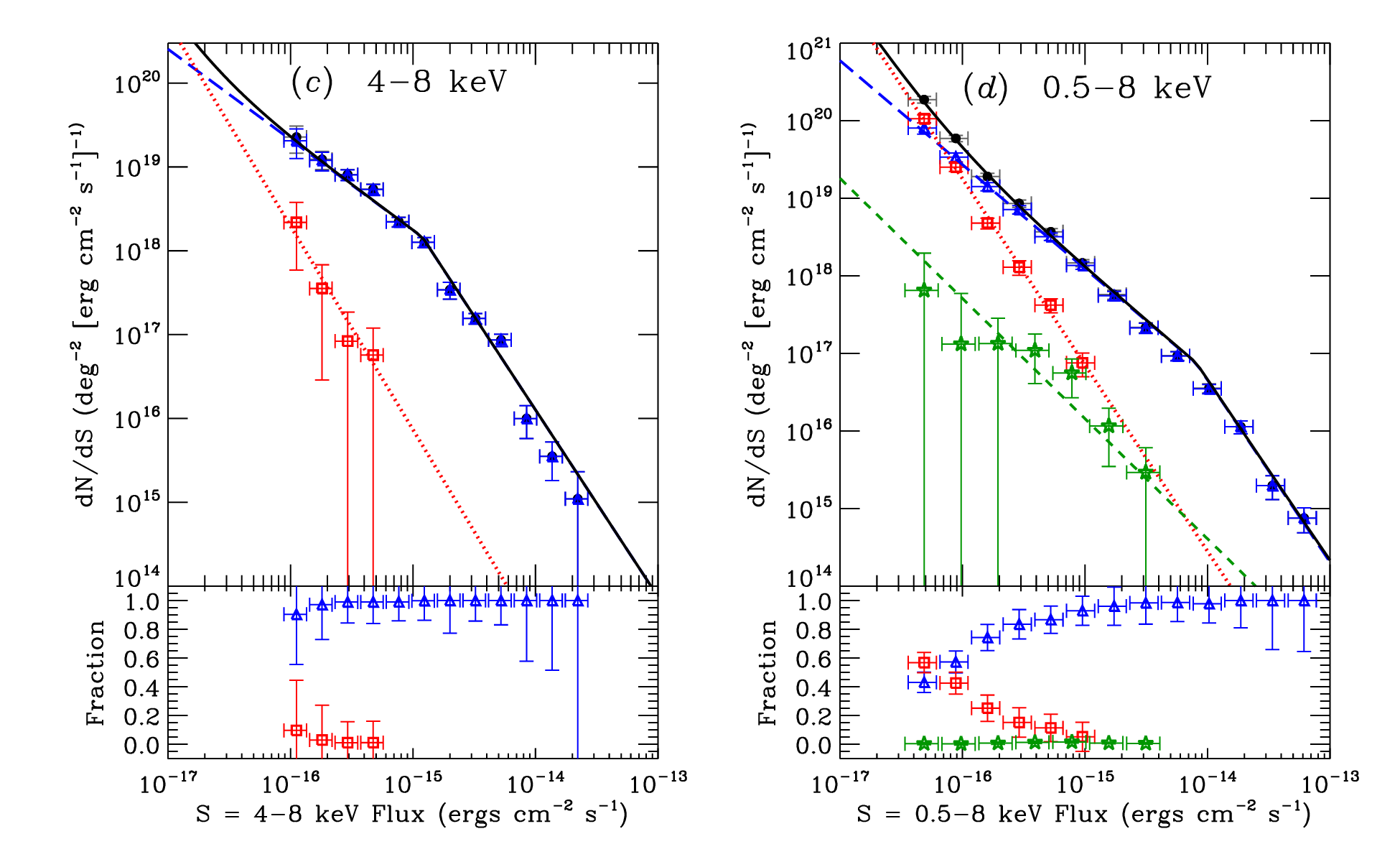}
}
\caption{
Differential number counts ($dN/dS$) versus flux for the SB ($a$), HB ($b$),
UHB ($c$), and FB ($d$), broken down into AGNs ({\it open blue triangles\/}),
normal galaxies ({\it open red squares\/}), and Galactic stars ({\it open green
stars\/}; for the SB, HB, and FB).  The differential number counts have been estimated using flux bins of $\approx$0.3~dex.
In each plot, the best-fit $dN/dS$
parameterizations based on equation~5 have been shown as blue long-dashed, red
dotted, and green short-dashed curves for AGNs, normal galaxies, and Galactic
stars, respectively.  The total number counts model, based on summing the three
contributing components, has been shown as a black curve.
{\it Bottom Panels of a--d}: Fractional contributions from AGNs, normal
galaxies, and Galactic stars to the differential number counts.  In the SB and
FB, we find that the differential number counts of normal galaxies appear to
surpass those of AGNs for fluxes below \hbox{$\approx$(1--2)~$\times
10^{-17}$~\flux} and \hbox{$\approx$(3--7)~$\times 10^{-17}$~\flux},
respectively.
}
\end{figure*}

\subsection{Cumulative Number Counts Computation}

Using the above information, we computed cumulative number counts for the
$\approx$4~Ms CDF-S survey using the following integral:
\begin{equation}
N(>S) = \int_{S}^{\infty} \left[ \displaystyle\sum\limits_{i=1}^{N_{\rm src}}
P_{T,i}(S)C^\prime_i/A(S, \Gamma_i) \right] dS,
\end{equation}
where the summation is over all \xray\ detected sources in a particular
category of sources (e.g., all sources, AGNs, normal galaxies, etc.) and
$\Gamma_i$ represents the effective photon index for source $i$.  For the 332
\xray\ sources detected in both the SB and HB, we utilized the photon indices
from X11; otherwise, we used $\Gamma = 1.4$.  Since $P_{{\rm T},i}$ is
dependent on our input Bayesian prior (which depends on source type), the
number counts will depend mildly on the choice of our model.  The term
$C^\prime$ is a count-rate and off-axis angle dependent completeness term,
which corrects for the fact that our \xray\ point-source catalogs were
constructed by (1) running {\ttfamily wavdetect} to form an initial list of
candidate sources and (2) assessing the probability of detection using
{\ttfamily AE}.  If we chose to use a catalog based strictly on the {\ttfamily
AE} source-detection probability assessment for every location on the image,
then $C^\prime = 1$ for all sources; however, such a catalog would produce
large numbers of false sources, leading to significant errors in the number
counts at faint fluxes, and is computationally impractical.  In practice, the
completeness corrections are small perturbations (median $C^\prime \approx
1.00$ for all bands, with $\approx$95\% of sources having $C^\prime
\simlt$~\hbox{1.7--2.4}) and affect number counts near the flux limits.
$C^\prime$ and its count-rate and off-axis angle dependencies were measured
using {\ttfamily marx} simulations, and a full description of this procedure
can be found in the appendix.  

%
%
\begin{figure*}
\figurenum{7}
\centerline{
\includegraphics[width=16cm]{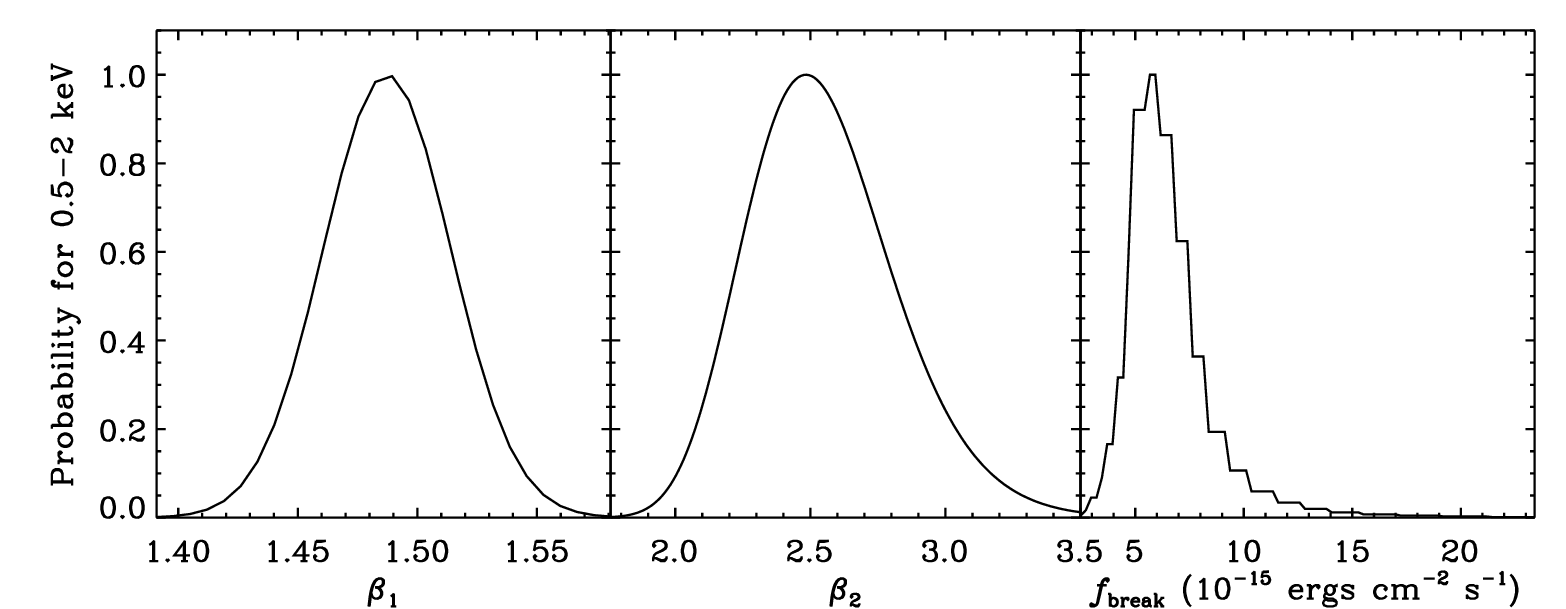}
}
\caption{
Sample Bayesian $dN/dS$ model parameter ($\beta_1^{\rm AGN}$, $\beta_2^{\rm
AGN}$, and $f_{\rm b}^{\rm AGN}$) likelihood distributions (normalized to the
maximum likelihood) for the \hbox{0.5--2~keV} band number counts.  These
parameters have been summarized in equation~5 and their
values are provided in Table~1.
}
\end{figure*}


\begin{table*}
\begin{center}
\caption{Maximum-Likelihood Best-Fit Model Parameters}
\begin{tabular}{lcccccccc}
\hline\hline
   & \multicolumn{4}{c}{AGNs} & \multicolumn{2}{c}{Normal Galaxies} & \multicolumn{2}{c}{Galactic Stars} \\
   & \multicolumn{4}{c}{\rule{2.5in}{0.01in}} & \multicolumn{2}{c}{\rule{1.5in}{0.01in}}  &\multicolumn{2}{c}{\rule{1.5in}{0.01in}} \\
 \multicolumn{1}{c}{Bandpass}  & $K^{\rm AGN}_{14}$ & $\beta_1^{\rm AGN}$  & $\beta_2^{\rm AGN}$ & $f_{\rm break}^{\rm AGN}$ & $K^{\rm gal}_{14}$ & $\beta^{\rm gal}$ & $K^{\rm star}_{14}$  & $\beta^{\rm star}$  \\
 \multicolumn{1}{c}{(1)} & (2) & (3) & (4) & (5) & (6) & (7) & (8) & (9) \\
\hline\hline
0.5--2~keV\ldots\ldots\ldots\ldots\ldots & 169.56~$\pm$~8.69 & $1.49 \pm 0.03$ & $2.48 \pm 0.27$ & 6.0$^{+1.4}_{-1.6}$ & 1.53~$\pm$~0.10 & 2.22$^{+0.08}_{-0.07}$&3.97~$\pm$~0.37 & 1.41$^{+0.14}_{-0.13}$\\
2--8~keV\dotfill & 573.13~$\pm$~27.49 & $1.32 \pm 0.04$ & 2.55$^{+0.17}_{-0.18}$ & $6.4 \pm 1.0$ & 1.10~$\pm$~0.22 & $2.29 \pm 0.25$&0.64~$\pm$~0.22 & $1.79 \pm 0.50$\\
4--8~keV\dotfill & 1463.27~$\pm$~63.61 & $1.08 \pm 0.10$ & $2.23 \pm 0.09$ & 1.2$^{+0.7}_{-0.8}$ & 0.27~$\pm$~0.10 & 2.43$^{+0.38}_{-0.40}$&\ldots & \ldots\\
0.5--8~keV\dotfill & 562.20~$\pm$~22.96 & 1.34$^{+0.04}_{-0.03}$ & $2.35 \pm 0.15$ & 8.1$^{+1.5}_{-1.4}$ & 2.82~$\pm$~0.26 & 2.40$^{+0.11}_{-0.12}$&4.07~$\pm$~0.51 & 1.55$^{+0.19}_{-0.18}$\\
\hline
\end{tabular}
\end{center}
NOTE.---Best-fit values and 1$\sigma$ errors for our number-counts priors.  Column~(1) lists the bandpass.  Columns~(2)--(5) provide the double power-law differential number-counts parameterization for AGNs, including the normalization $K_{\rm 14}^{\rm AGN}$ (Col.~2; in units of $10^{14}$~deg$^{-2}$~[\flux]$^{-1}$), faint-end slope $\beta_1^{\rm AGN}$ (Col.~3), bright-end slope $\beta_2^{\rm AGN}$ (Col.~4), and break-flux $f_{\rm break}^{\rm AGN}$ (Col.~5; in units of $10^{-15}$ \flux).  Columns~(6) and (7) provide the normal-galaxy single-power-law normalization $K_{\rm 14}^{\rm gal}$ and slope $\beta^{\rm gal}$, respectively.  Columns~(8) and (9) provide the Galactic star single-power-law normalization $K_{\rm 14}^{\rm star}$ and slope $\beta^{\rm star}$, respectively.  These values were computed using the maximum-likelihood methods described in $\S$2.2.
\end{table*}

Following the same approach implemented in $\S$5 of G08, we utilized
maximum-likelihood techniques to optimize the Bayesian model parameters.  As
discussed above, we made use of power-law differential number-counts models for
our Bayesian priors (see equation~5), which are dependent on source type.  For
a given source, we first characterized its type (i.e., AGN, normal galaxy, or
Galactic star; see $\S$3.1 below), chose an appropriate model in equation~5,
and computed the probability $p_i$ of that source being present in the
$\approx$4~Ms \cdfs\ catalog following
\begin{equation}
p_{\rm i} = \frac{\int P_T(S, N_i) dS}{\int dN/dS \big|_{{\rm model},i}
A(S,\Gamma_i) dS},
\end{equation}
where $N_i$ represents the total number of counts measured within the
extraction cell for each detected source.  Considering our model, the total
likelihood of obtaining the $\approx$4~Ms source catalog and its source-count
distribution can be computed as $\prod_{\rm i} p_i$.  We maximized the total
likelihood for the model input parameters for AGNs, normal galaxies, and
Galactic stars separately to find best-fit values for each of the parameters in
equation~5.

%
\section{Results}
%

\subsection{Number Counts by Source Type and Total Counts}

The extensive multiwavelength data in the $\approx$4~Ms \cdfs\ region allow for
the robust characterization of the \xray\ detected sources.  As described in
X11, 716 of the 740 \xray\ detected sources have multiwavelength counterparts,
and 673 have either secure spectroscopic or reliable photometric redshifts.
Following a similar scheme to that provided by X11, we have classified the
\xray\ sources in our sample as active galactic nuclei (AGNs), normal
star-forming galaxies (hereafter, galaxies), and Galactic stars.  In this
process, we make use of six AGN selection criteria:
\begin{enumerate}
\item {\itshape X-ray Luminosity}: A source with an observed-frame SB, HB, UHB,
or FB luminosity (i.e., observed-frame $L_{\rm X} = 4\pi d_L^2 f_{\rm X}$;
where $d_L$ is the luminosity distance) $\simgt 3 \times 10^{42}$~\lum\ is
classified as a luminous AGN.  In the nearby universe, the most powerful
star-forming galaxies (luminous infrared galaxies) all have $L_{\rm 0.5-8~keV}
\simlt 10^{42}$~\lum\ (e.g., Iwasawa \etal\ 2009, 2011; Lehmer \etal\ 2010;
Pereira-Santaella \etal\ 2011).
\item {\itshape X-ray Spectral Shape}: A source with an effective photon index
of $\Gamma \simlt 1.0$ is most likely to be dominated by a single powerful
\xray\ source that is significantly obscured.  We consider such sources to be
obscured AGNs.
\item {\itshape X-ray to Optical Flux Ratio}: A source with an \xray\ to
optical flux ratio of $\log (f_{\rm X}/f_R) > −1$ (where $f_{\rm X} = f_{\rm
0.5-2~keV}$, $f_{\rm 2-8~keV}$, $f_{\rm 4-8~keV}$, or $f_{\rm 0.5-8~keV}$) is
considered to have \xray\ emission significantly elevated compared with
normal-galaxy processes (e.g., hot gas, low-mass and high-mass \xray\ binary
emission) as traced by stellar emission (i.e., $f_R$).  Such sources were
classified as AGNs.
\item {\itshape X-ray to Radio Luminosity Ratio}: A source with substantial
excess (i.e., a factor of $\simgt$5) X-ray emission over the level expected
from pure star formation (i.e., observed-frame SB, HB, FB, or UHB luminosity
$\simgt 5 \times [8.9 \times 10^{17} L_{\rm 1.4~GHz}]$; Bauer \etal\ 2002) is
considered to be \xray\ overluminous compared with the \xray/SFR correlation
and its \xray\ emission is therefore likely to be dominated by an AGN.  This
criterion is similar in nature to that of criterion~3, but differs in that it
allows for a more sensitive classification of a subset of actively star-forming
galaxies (i.e., detected at 1.4~GHz) using a more reliable tracer of the
intrinsic galactic SFR that is insensitive to dust obscuration.
\item {\itshape X-ray Variability}: Sources found to be variable in the \xray\
band (with probability $\simgt$95\%) on timescales of months to years that also
have \hbox{0.5--8~keV} luminosities greater than $10^{41}$~\lum\ are unlikely
to be produced by normal-galaxy processes (see Young \etal\ 2012 for details).
These sources were classified as AGNs.  
\item {\itshape Optical Spectroscopy}: Finally, sources with optical
spectroscopic AGN features such as broad and/or high-excitation emission lines
are classified as AGNs.  These sources were identified using the spectroscopic
catalogs of Szokoly \etal\ (2004), Mignoli \etal\ (2005), and Silverman \etal\
(2010) (see X11 for details).
\end{enumerate} 

%
%
\begin{figure*} \figurenum{8} \centerline{
\includegraphics[width=16cm]{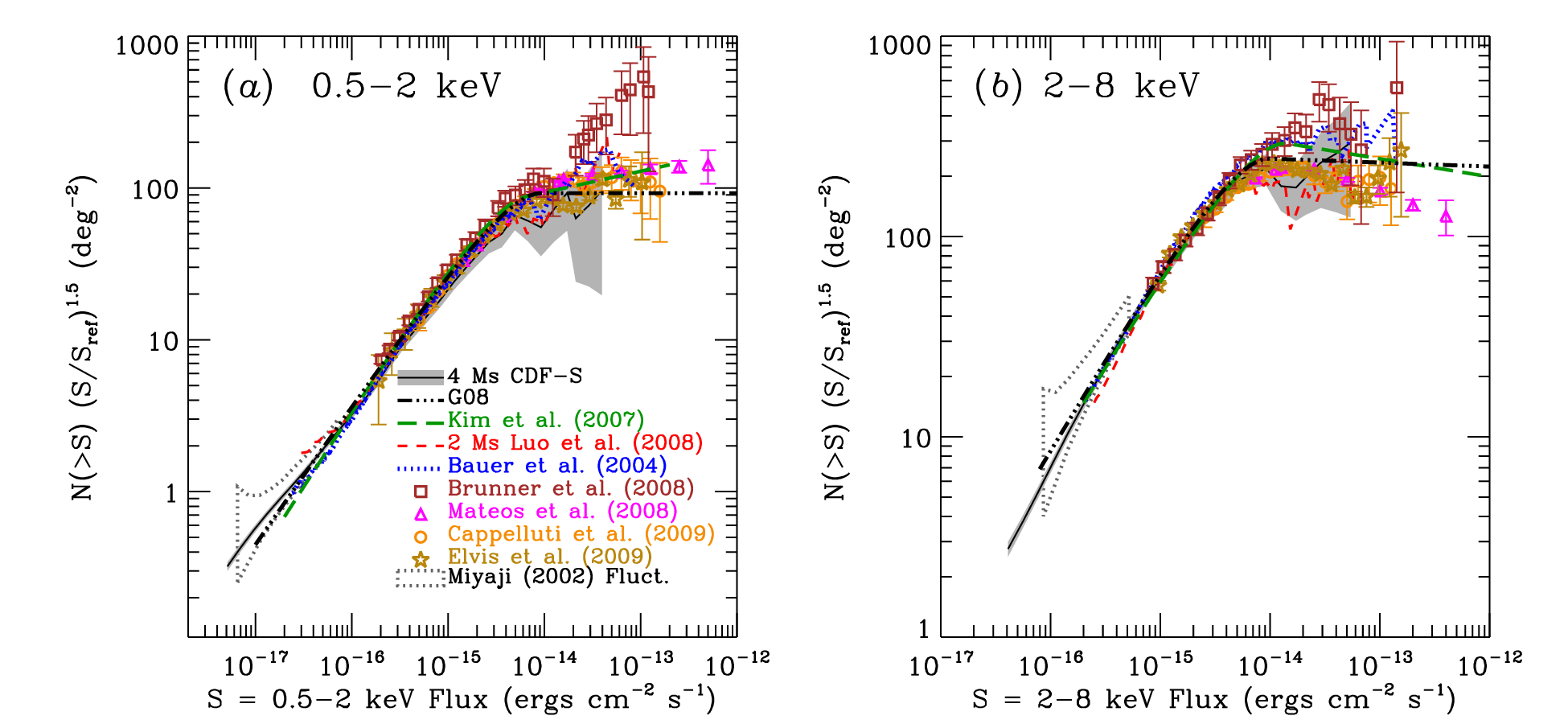} } \centerline{
\includegraphics[width=16cm]{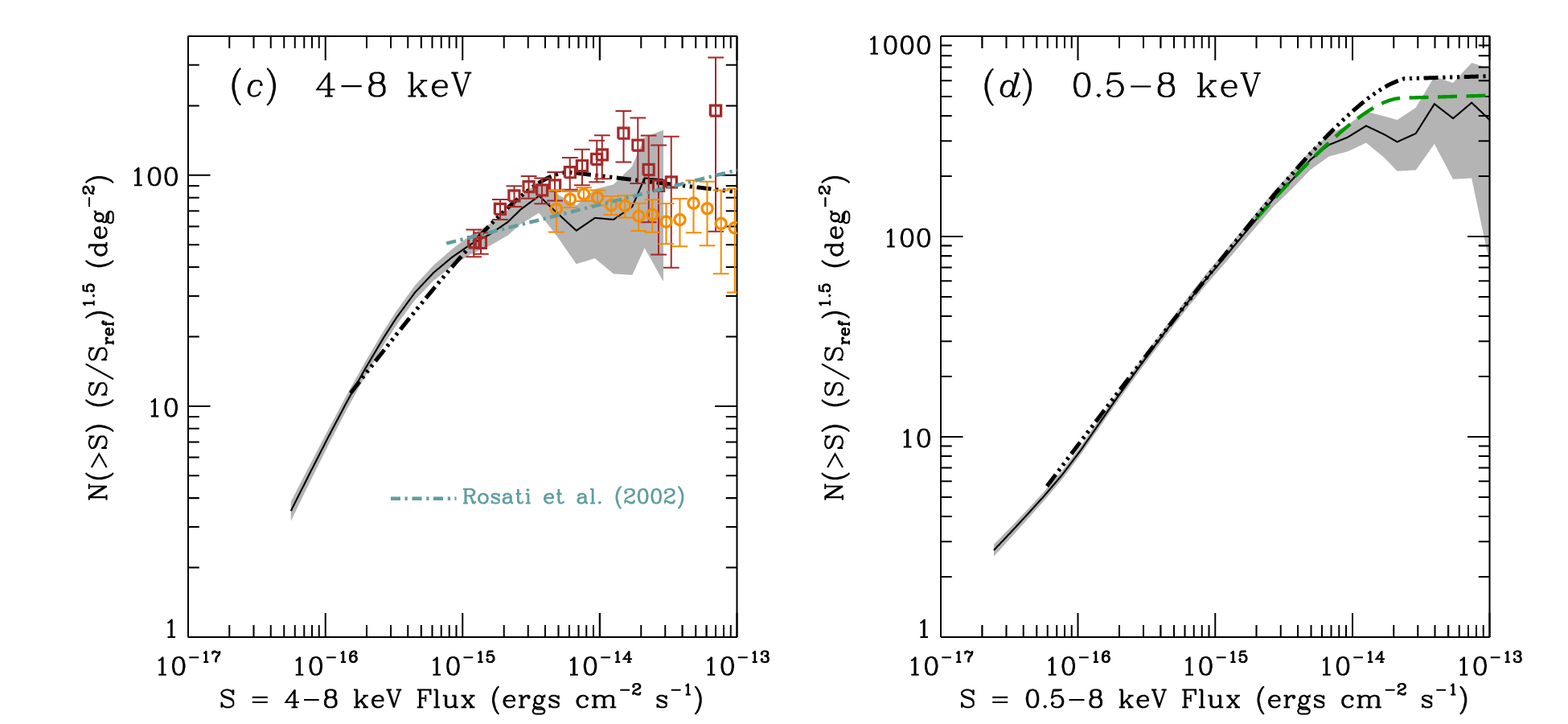} } \caption{
Euclidean-normalized, cumulative number counts ($N[>S] \times [S/S_{\rm
ref}]^{1.5}$, where $S_{\rm ref} = 10^{-14}$~\flux) and 1$\sigma$ errors for
the $\approx$4~Ms CDF-S ({\it black curves with gray error bars\/}) for the SB
({\it a}), HB ({\it b\/}), UHB ({\it c\/}), and FB ({\it d\/}).  The Miyaji \&
Griffiths (2002) predicted number-counts boundaries (from $\approx$1~Ms
fluctuation analyses) are shown as the dotted regions in panels $a$ and $b$.
Previous $\approx$2~Ms \cdfn\ plus $\approx$1~Ms \cdfs\ number counts estimates
from B04 and $\approx$2~Ms \cdfs\ number counts from L08 are indicated as blue
dotted and red short-dashed curves, respectively.  The G08 number counts, which
include the $\approx$2~Ms \cdfn\ and $\approx$1~Ms \cdfs\ surveys, are
indicated as dashed triple-dotted curves.  Similarly, the ChaMP survey number
counts from Kim \etal\ (2007), which include the $\approx$2~Ms \cdfn\ and
$\approx$1~Ms \cdfs\ surveys, have been shown as green long-dashed curves.
Number-counts measurements from other \xmm\ and \chandra\ surveys have been
indicated, including the \xmm\ serendipitous (Mateos \etal\ 2008; {\it magenta
triangles\/}), Lockman hole (Brunner \etal\ 2008; {\it brown squares\/}), and
COSMOS (Cappelluti \etal\ 2009; {\it orange circles\/}) surveys, as well as the
\chandra\ COSMOS (Elvis \etal\ 2009; see also Puccetti \etal\ 2009; {\it gold
stars\/}).  In the case of the UHB, we highlight the $\approx$1~Ms \cdfs\
number counts from Rosati \etal\ (2002; {\it cyan dot-dashed curve\/}).  We
note that all measurements shown (i.e., both {\it curves} and {\it data
points\/}) correspond to measurements, and models are {\it not} plotted here.
We find good overall agreement between our number counts measurements and those
found in previous studies.
[{\it A machine-readable table of the cumulative number counts data for the $\approx$4~Ms CDF-S is
provided in the electronic edition.}]
} \end{figure*}

%

\begin{sidewaystable*}[p]
\vspace{2in}
\begin{center}
\caption{Number Counts Statistics}
\begin{tabular}{llcccccccccccc}
\hline\hline
   &  & & & & & \multicolumn{4}{c}{{\sc Total}} & \multicolumn{4}{c}{{\sc CXRB Fraction}}  \\
   &  & \multicolumn{4}{c}{{\sc Number}} & \multicolumn{4}{c}{(deg$^{-2}$)} & \multicolumn{4}{c}{(\%)}  \\
   &  & \multicolumn{4}{c}{\rule{1.0in}{0.01in}} & \multicolumn{4}{c}{\rule{2.4in}{0.01in}} & \multicolumn{4}{c}{\rule{2.4in}{0.01in}}  \\
 \multicolumn{1}{c}{\sc Class}  & \multicolumn{1}{c}{\sc Subclass}  & SB & HB & UHB & FB &SB & HB & UHB & FB &SB & HB & UHB & FB \\
 \multicolumn{1}{c}{(1)}  & \multicolumn{1}{c}{(2)}  & {(3)}  & {(4)}  & {(5)}  & {(6)}  & {(7)}  & {(8)} & (9) & (10) & (11) & (12) & (13) & (14) \\
\hline\hline
All + Bright-Correction$^a$\ldots\ldots\ldots\ldots\dotfill & Total & \ldots  & \ldots  & \ldots & \ldots  & \ldots & \ldots & \ldots & \ldots & 75.7 $\pm$ 4.3 & 82.4 $\pm$ 13.0 & 88.4 $\pm$ 13.8 & 81.6 $\pm$ 8.9 \\ 
All\dotfill & Total & 650 & 403 & 260 & 634 & 27832 $\pm$ 1803 & 10495 $\pm$ 871 & 8387 $\pm$ 787 & 22579 $\pm$ 1506 & 53.1 $\pm$ 3.4 & 66.3 $\pm$ 5.8 & 64.6 $\pm$ 6.7 & 62.7 $\pm$ 4.2 \\ 
AGNs \dotfill & Total & 474 & 387 & 256 & 520 & 14925 $\pm$ 1228 & 9310 $\pm$ 776 & 8053 $\pm$ 757 & 12802 $\pm$ 943 & 49.0 $\pm$ 4.0 & 65.5 $\pm$ 5.8 & 64.1 $\pm$ 6.6 & 59.6 $\pm$ 4.4 \\ 
\\
 & unknown $z$ & 43 & 21 & 9 & 42 & 2603 $\pm$ 545 & 1058 $\pm$ 326 & 622 $\pm$ 249 & 1906 $\pm$ 412 & 1.1 $\pm$ 0.2 & 1.9 $\pm$ 0.6 & 1.9 $\pm$ 0.8 & 1.8 $\pm$ 0.4 \\ 
\\
 & $z = $~0.0--1.5 & 209 & 182 & 123 & 238 & 5630 $\pm$ 721 & 3929 $\pm$ 478 & 3677 $\pm$ 537 & 5077 $\pm$ 523 & 31.6 $\pm$ 4.0 & 40.0 $\pm$ 5.2 & 37.9 $\pm$ 6.1 & 37.3 $\pm$ 3.8 \\ 
 & $z = $~1.5--3.0 & 165 & 133 & 89 & 181 & 5338 $\pm$ 763 & 3111 $\pm$ 428 & 2780 $\pm$ 413 & 4738 $\pm$ 612 & 12.9 $\pm$ 1.8 & 17.7 $\pm$ 2.6 & 17.3 $\pm$ 2.8 & 15.9 $\pm$ 2.1 \\ 
 & $z >$~3.0 & 57 & 51 & 35 & 59 & 1353 $\pm$ 328 & 1213 $\pm$ 289 & 975 $\pm$ 228 & 1082 $\pm$ 266 & 3.4 $\pm$ 0.8 & 5.9 $\pm$ 1.5 & 7.1 $\pm$ 1.8 & 4.7 $\pm$ 1.1 \\ 
\\
 & $\log N_{\rm H} <$~22.0 & 102 & 86 & 51 & 108 & 993 $\pm$ 105 & 1559 $\pm$ 262 & 828 $\pm$ 159 & 1572 $\pm$ 40 & 30.3 $\pm$ 3.2 & 21.5 $\pm$ 3.9 &16.1 $\pm$ 3.4 & 24.7 $\pm$ 0.6 \\
 & $\log N_{\rm H} = $~22.0--23.0 & 224 & 177 & 115 & 249 & 6871 $\pm$ 818 & 4589 $\pm$ 558 & 4157 $\pm$ 583 & 6548 $\pm$ 81 & 14.4 $\pm$ 1.7 & 29.0 $\pm$ 3.7 &28.1 $\pm$ 4.3 & 24.3 $\pm$ 0.3 \\
 & $\log N_{\rm H} >$~23.0 & 105 & 103 & 81 & 121 & 4458 $\pm$ 728 & 2105 $\pm$ 339 & 2447 $\pm$ 382 & 2777 $\pm$ 54 & 3.1 $\pm$ 0.5 & 13.0 $\pm$ 2.2 &18.0 $\pm$ 3.0 & 8.9 $\pm$ 0.2 \\
\\
 & $\log L_{\rm X} <$~42.0 & 44 & 23 & 12 & 49 & 2283 $\pm$ 544 & 889 $\pm$ 243 & 803 $\pm$ 304 & 2111 $\pm$ 423 & 1.6 $\pm$ 0.4 & 1.7 $\pm$ 0.5 & 2.0 $\pm$ 0.8 & 1.9 $\pm$ 0.4 \\ 
 & $\log L_{\rm X} = $~42.0--43.0 & 124 & 86 & 40 & 142 & 5086 $\pm$ 704 & 3247 $\pm$ 514 & 1592 $\pm$ 359 & 5348 $\pm$ 677 & 3.4 $\pm$ 0.5 & 6.3 $\pm$ 1.0 & 6.1 $\pm$ 1.5 & 5.3 $\pm$ 0.7 \\ 
 & $\log L_{\rm X} = $~43.0--44.0 & 167 & 159 & 113 & 186 & 3818 $\pm$ 626 & 3067 $\pm$ 388 & 3611 $\pm$ 485 & 2534 $\pm$ 267 & 12.4 $\pm$ 2.0 & 21.6 $\pm$ 2.9 & 24.4 $\pm$ 3.6 & 17.9 $\pm$ 1.9 \\ 
 & $\log L_{\rm X} >$~44.0 & 96 & 98 & 82 & 101 & 1134 $\pm$ 167 & 1050 $\pm$ 149 & 1425 $\pm$ 234 & 904 $\pm$ 102 & 30.4 $\pm$ 4.5 & 34.0 $\pm$ 5.2 & 29.8 $\pm$ 5.4 & 32.7 $\pm$ 3.7 \\ 
\\
Normal Galaxies \dotfill & Total & 166 & 14 & 4 & 106 & 12704 $\pm$ 1316 & 1142 $\pm$ 394 & 334 $\pm$ 216 & 9638 $\pm$ 1172 & 3.0 $\pm$ 0.3 & 0.8 $\pm$ 0.3 & 0.5 $\pm$ 0.3 & 2.6 $\pm$ 0.3 \\ 
\\
 &  $z \simlt 0.6$ & 82 & 11 & 3 & 63 & 4924 $\pm$ 792 & 975 $\pm$ 371 & 165 $\pm$ 155 & 5049 $\pm$ 825 & 1.6 $\pm$ 0.2 & 0.7 $\pm$ 0.2 & 0.3 $\pm$ 0.3 & 1.6 $\pm$ 0.3 \\ 
 &  $z \simgt$~0.6 & 84 & 3 & 1 & 43 & 7770 $\pm$ 1052 & 167 $\pm$ 133 &  \ldots & 4573 $\pm$ 831 & 1.4 $\pm$ 0.2 & 0.1 $\pm$ 0.1 &  \ldots & 1.0 $\pm$ 0.2 \\ 
\\
 &  Late-Type (star-forming)& 123 & 10 & 1 & 74 & 10174 $\pm$ 1191 & 1051 $\pm$ 390 &  \ldots & 7222 $\pm$ 1014 & 2.2 $\pm$ 0.3 & 0.6 $\pm$ 0.2 &  \ldots & 1.8 $\pm$ 0.3 \\ 
 &  Early-Type (passive) & 43 & 4 & 3 & 32 & 2530 $\pm$ 561 & 91 $\pm$ 53 & 294 $\pm$ 210 & 2416 $\pm$ 588 & 0.8 $\pm$ 0.2 & 0.2 $\pm$ 0.1 & 0.4 $\pm$ 0.3 & 0.8 $\pm$ 0.2 \\ 
\\
Stars \dotfill & Total & 10 & 2 & \ldots & 8 & 202 $\pm$ 109 & 34 $\pm$ 32 &  \ldots &139 $\pm$ 88 & 1.2 $\pm$ 0.6 & 0.1 $\pm$ 0.1 &  \ldots  & 0.5 $\pm$ 0.3 \\ 
\hline
\end{tabular}
\end{center}
$^a$CXRB fractions have been computed using the number-counts measurements presented in this paper below $5 \times 10^{-15}$, $1.4 \times 10^{-14}$, $8.4 \times 10^{-15}$, and $2 \times 10^{-15}$~\flux\ for the SB, HB, UHB, and FB, respectively, plus bright-end corrections from the Kim \etal\ (2007) number-counts relations (see $\S$~3.4 for details).\\
NOTE.---Number-counts statistics for the source classifications (Column~1) and subclassifications (Column~2) discussed in this paper.  Columns~(3)--(6) provide the number of sources in each subclass used in calculating number counts for the four bandpasses.  Columns~(7)--(10) provide the number counts for each subcategory at the limiting fluxes for the SB (\sblim~\flux), HB (\hblim~\flux), UHB (\uhblim~\flux), and FB (\fblim~\flux), respectively. Columns~(11)--(14) provide the integrated intensity (expressed as the percentage of the CXRB intensity) for the SB, HB, UHB, and FB, assuming CXRB intensity values of \sbcxrb, \hbcxrb, \uhbcxrb, and \fbcxrb~\intensity, respectively (see $\S$~3.4 for details).
\end{sidewaystable*}

A source is initially classified as an AGN if at least one of the above six
criteria are satisfied.  However, we found that five of the sources that were
classified as AGNs via the above criteria had \hbox{1--2~keV} and
\hbox{0.5--1~keV} band ratios that were consistent with being dominated by hot
interstellar gas (see Danielson \etal\ 2012).  The initial AGN classification
for these five sources was based on having high \xray\ to optical flux ratios
(criterion~3); however, their \xray\ luminosities compared with their $B$-band
luminosities appear to be consistent with hot-gas dominated galaxies (see
Danielson \etal\ 2012 for details).  We therefore classified these five objects
as normal galaxies.  In addition to classifying sources directly as AGNs and
normal galaxies, we searched for direct multiwavelength indicators of Galactic
stars.  Likely Galactic stars were identified using (1) optical spectroscopy
from Szokoly \etal\ (2004), Mignoli \etal\ (2005), and Silverman \etal\ (2010);
(2) \hst\ stellarity indices $>$0.7 (from the Caldwell \etal\ 2008 GEMS
catalogs); and (3) best-fit stellar templates from MUSYC photometric-redshift
fits (Cardamone \etal\ 2010).  All Galactic star candidates were visually
screened and obvious normal galaxies were noted.  The remaining stellar
candidates were then classified as ``Galactic stars.'' Remaining sources that
were not classified as either AGNs or stars were classified as normal galaxies.
Based on these criteria, we therefore estimate that of the 740 \xray\ detected
sources, 561 are AGNs, 169 are normal galaxies, and 10 are Galactic stars.  

We compared our normal-galaxy classifications with the 31 radio and \xray\
selected normal galaxies classified by Vattakunnel \etal\ (2012) and found that
25 sources overlap.  We find that 4 of the Vattakunnel \etal\ (2012) sources
that we classify as AGNs had high X-ray to radio luminosity ratios
(criterion~4), a criterion that was not implemented by Vattakunnel \etal\ (2012).

In Figure~4, we plot the observed-frame \hbox{0.5--8~keV} luminosity versus
redshift for the AGNs and normal galaxies with redshift measurements.  Our AGN
and normal galaxy samples span redshift ranges of $z \approx$~\hbox{0.1--8} and
$z \approx$~\hbox{0.03--2.6}, respectively.  These redshifts are from both
photometric and spectroscopic redshift estimates, and all redshift estimates
above $z = 4.76$ are based on photometric redshifts.  We have chosen to use the
most probable photometric redshift for each source; however, some of the $z >
4.76$ sources have photometric redshifts consistent with being at lower
redshifts (see Luo \etal\ 2010 for details).

Using the classifications adopted above and our methods for computing and
optimizing number counts described in $\S$2, we computed the cumulative number
counts for AGNs, normal galaxies, and Galactic stars, and determined the
best-fit parameters related to their priors (i.e., those characterized in
equation~5).  In Figure~5, we present the breakdown of the cumulative number
counts from AGNs, normal galaxies, and Galactic stars for the four bandpasses
({\it discrete symbols\/}).  Total number counts for the survey are simply the
sum of the number counts from each of these populations.  In Figure~6, we show
the {\it differential} number counts ($dN/dS$) for the
four bandpasses.  The best-fit $dN/dS$ model parameters, used in computing our
number counts, have been tabulated in Table~1, and the cumulative and
differential number counts corresponding to these models have been shown in
Figures~5 and 6, respectively ({\it continuous curves\/}).  In Figure~7, we
show sample probability distributions for the best-fit $dN/dS$ model parameters
appropriate for AGNs detected in the SB.  

%
%
\begin{figure*}
\figurenum{9}
\centerline{
\includegraphics[width=16cm]{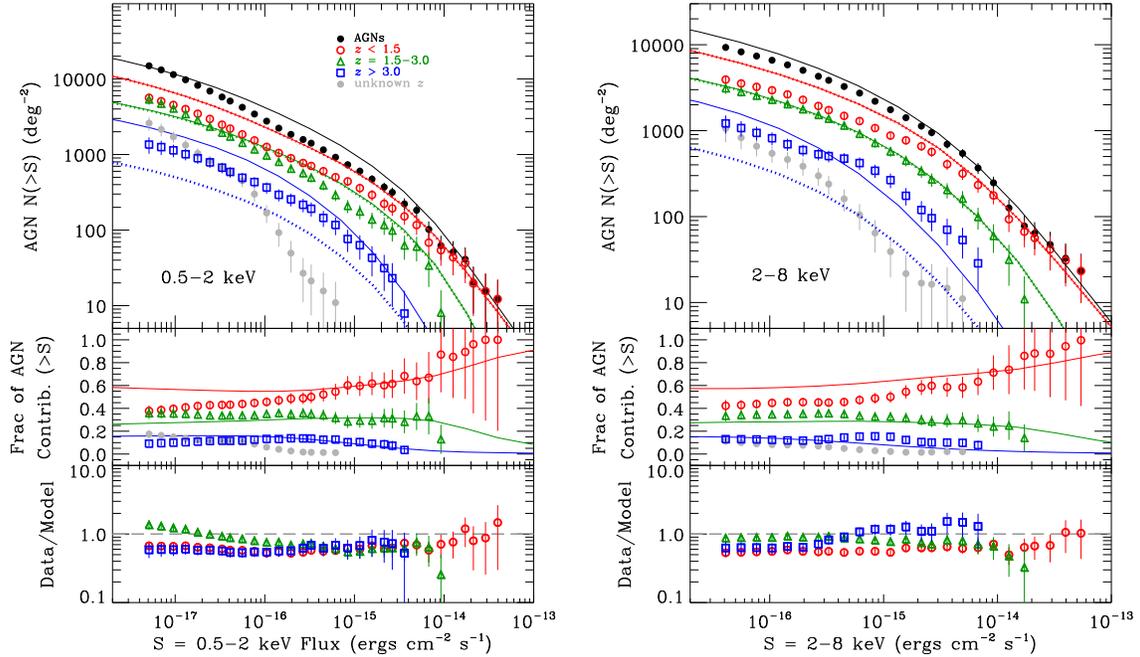}
}
\caption{
{\it Top Panels}: Cumulative AGN number counts ({\it filled black circles\/})
in the SB ({\it left\/}) and HB ({\it right\/}) broken down by contributions
from AGNs in redshift ranges of $z < 1.5$ ({\it open red circles\/}), $z
=$~1.5--3 ({\it open green triangles\/}), $z > 3$ ({\it open blue squares\/}),
and no measured redshift ({\it filled gray circles\/}).  AGN number-count
predictions from the AGN population-synthesis models by G07 have been provided
both with ({\it dotted curves\/}) and without ({\it solid curves\/}) an
exponential decline in the luminosity function at $z > 2.7$.  G07 AGN
number-counts models for all redshifts, $z < 1.5$, $z =$~1.5--3, and $z > 3$
have been shown as black, red, green, and blue curves, respectively.  These
models are discussed in detail in $\S$~4.1.
{\it Middle Panels}: Fractional contribution that AGNs in each redshift range
provide to the overall AGN number counts. The solid curves are the expectations
from the G07 model without a declining XLF at $z > 2.7$.
{\it Bottom Panels}: The ratio of the AGN number-counts data presented in the
top panel divided by the G07 model (without a declining XLF at $z > 2.7$).
[{\it A machine-readable table of the redshift-divided AGN cumulative number
counts data for all four bandpasses is provided in the electronic edition.}]
}
\end{figure*}

The bottom panels of Figures~5 and 6 show the fractional contribution that the
respective cumulative and differential number counts of each source type make
to the total counts.  We find that AGNs still dominate the cumulative number
counts at all flux levels in all four bands; however, at SB and FB fluxes below
$\simlt 10^{-16}$~\flux, normal galaxies undergo a rapid increase in their
numbers.  At the faintest SB flux limit ($\approx$\sblim~\flux), AGNs and
normal galaxies reach source densities of $\approx$\agnden\ and
\galden~deg$^{-2}$, respectively.  These sky densities are factors of
$\approx$2.2 and $\approx$4.5 times larger than those reported at the
$\approx$2~Ms \chandra\ depth (see B04) for AGNs and normal galaxies,
respectively.   We find that our SB normal galaxy number counts lie between
the ``optimistic'' and ``pessimistic'' estimates from B04 and are in good agreement
with previous work by Hornschemeier \etal\ (2003).  Due to the small solid
angle of our survey, the bright normal galaxy number counts data drop off
at SB fluxes $\simgt$$4\times 10^{-16}$~\flux; however, the extension of our best-fit
power-law fit ({\it red dotted curve} in Figure~5$a$) agrees well with
bright-end normal galaxy number counts from the \xmm\ serendipitous plus
needles in the haystack surveys (Georgakakis \etal\ 2006) and the \chandra\
Multiwavelength Project (Kim \etal\ 2006).
Normal galaxies comprise 46 $\pm$ 5\% and 43 $\pm$ 5\% of the
total cumulative number counts in the SB and FB, respectively, a result
consistent with that found by X11, who showed that $\approx$40--50\% of the
SB-detected sources within $\approx$4\arcmin\ of the $\approx$4~Ms \cdfs\
center are normal galaxies (see their Fig.~13c).  This signifies more than a
doubling in the fractional contribution that the normal-galaxy cumulative
number counts make to the total number counts at the SB flux limit over that
found in the $\approx$2~Ms CDFs (see B04).  In terms of the differential number
counts, it appears that the contribution from normal galaxies exceeds that of
AGNs at SB and FB fluxes below $\approx$(1--2)~$\times 10^{-17}$~\flux\ and
$\approx$(3--7)~$\times 10^{-17}$~\flux, respectively (see Fig.~6$a$).  We
estimate that near the SB and FB flux limits, normal galaxies make up
\hbox{$\approx$45--70\%} of the differential number counts.  The rapid increase
in normal-galaxy counts signifies that deep \chandra\ surveys are on the verge
of being normal-galaxy dominated at the faintest fluxes (see $\S$5 for a
discussion).  In Table~2, we summarize properties of the number counts.  

In Figure~8, we show our derived total number-count distributions normalized by
a Euclidean slope of $S^{1.5}$ (i.e., \hbox{$N[>S] \times [S/S_{\rm
ref}]^{1.5}$}, where $S_{\rm ref} = 10^{-14}$~\flux).  The Euclidean
representation allows for comparisons with previous investigations at both the
faint and bright ends of the distributions.  For comparison, we have plotted
number-count distributions from previous surveys with \xmm\ (Brunner \etal\
2008; Mateos \etal\ 2008; Cappelluti \etal\ 2009) and \chandra\ (Bauer \etal\
2004; Kim \etal\ 2007; G08; Luo \etal\ 2008; Elvis \etal\ 2009; Puccetti \etal\
2009).
Some small differences in flux measurements (at the $\simlt$10\% level) are
expected due to both changes in calibration over the years for
\chandra,\footnote{For example, as noted in X11, in January, 2009 the \chandra\
ACIS-I ancillary response file (ARF) was updated resulting in a flat
$\approx$9\% reduction in effective area and a $\approx$0--8\% reduction
between \hbox{2--5~keV} (see
http://cxc.harvard.edu/ciao/why/caldb4.1.1\_hrma.html). Further uncertainties,
at the $\approx$10\% level, are expected due to spatial variations of
contamination on the optical-blocking filter (see
http://web.mit.edu/iachec/meetings/2011/Presentations/Marshall.pdf). } as well
as cross-calibration uncertainties between \chandra\ and \xmm\ (see, e.g.,
Nevalainen \etal\ 2010).  In general, we find good agreement between our number
counts and those of previous studies with some minor differences.  At faint SB
fluxes, we find that our number-counts measurements are in good agreement with
those of G08 and B04, which are both based on the $\approx$2~Ms \cdfn\ and
$\approx$1~Ms \cdfs; however, our estimates are \hbox{$\approx$30--40\%} lower
than those near the limit of the $\approx$2~Ms \cdfs\ (L08).  It is likely that
this discrepancy is due to an incompleteness overcorrection of the faint-end
counts in L08.  In the HB, we find that our number counts are in good agreement
with those of B04, but are somewhat lower than those of G08 and higher than
those of L08 (at the $\approx$10--20\% level in both cases).  In the UHB, we
find clear evidence of a break in the number counts at $f_{\rm 4-8~keV} \approx
10^{-15}$~\flux\ (see Figs.~5--6).  Previous investigations of the UHB in the
CDF-S revealed no obvious break down to $\approx$$7\times 10^{-16}$~\flux\
(Rosati \etal\ 2002), which is very close to where we observe the break.  We
note that the G08 investigation found a break at a much brighter flux of
$\approx$$6 \times 10^{-15}$~\flux.  In the CDF-S, there are only $\approx$17
UHB sources brighter than this flux, which would make it difficult to identify
such a break if present.  In contrast, the G08 analysis made use of wide-area
surveys like the EGS, ELAIS-N1, and XBOOTES, and therefore have better source
statistics at bright UHB fluxes.  In all four bands, the $\approx$4~Ms
\cdfs\ number counts measurements have large uncertainties at bright fluxes due
to the relatively small solid angle of the survey; however, we find good
general agreement between our measurements and those of previous
investigations.

In $\S$$\S$~3.2 and 3.3 below, we further divide our number-counts
measurements into contributions from AGN and normal-galaxy subpopulations.  For
ease of presentation, we discuss only results from the SB and HB; however, full
results from all four bandpasses are provided in the tables for reference.

%
%
\begin{figure}
\figurenum{10}
\centerline{
\includegraphics[width=9cm]{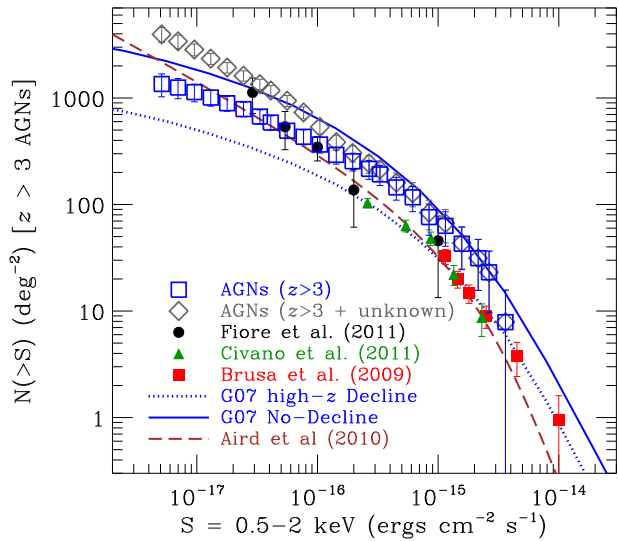}
}
\caption{
Cumulative SB AGN number counts for sources at redshifts $z \simgt 3$.  The
$\approx$4~Ms \cdfs\ estimates and 1$\sigma$ error bars have been shown as open
blue squares with error bars.  If we add to the $z \simgt 3$ AGN number counts
the number counts for \xray\ detected sources without known redshifts, we
obtain the open gray diamonds with error bars.  Due to their lack of
multiwavelength counterparts, in regions where extensive multiwavelength
coverage is available, the sources with unknown redshifts are good candidates
for being at $z \simgt 3$.  Recent constraints from Fiore \etal\ (2011) for the
$\approx$4~Ms \cdfs\ ({\it filled black circles\/}), Civano \etal\ (2011) for
\chandra\ COSMOS ({\it filled green triangles\/}), and Brusa \etal\ (2009) for
\xmm\ COSMOS ({\it filled red squares\/}) have been shown for comparison.  The
G07 models for $z \simgt 3$ number counts have been shown for the cases where a
high-redshift ($z > 2.7$) space-density decline is ({\it blue dotted curve\/})
and is not ({\it blue solid curve\/}) included.  The recent number-counts
predictions, based on the Aird \etal\ (2010) XLF measurement in the 2--10~keV
band, are shown as a brown long-dashed curve.
}
\end{figure}

%
%
\begin{figure*}
\figurenum{11}
\centerline{
\includegraphics[width=16cm]{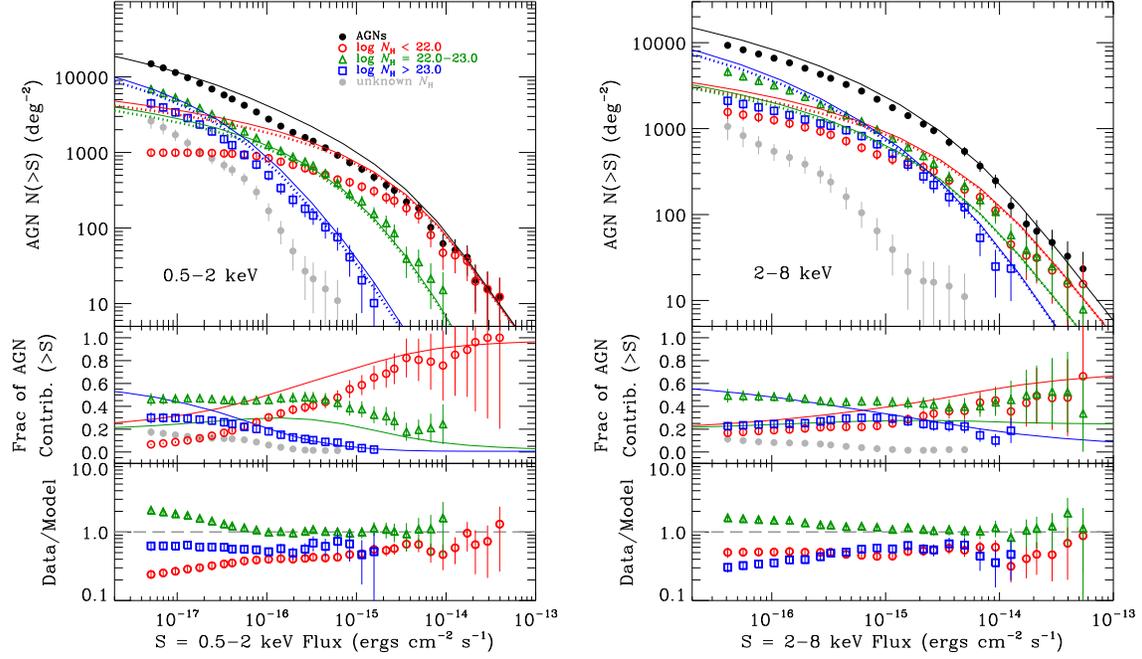}
}
\caption{
Similar to Figure~9, but for AGNs divided into ranges of intrinsic column density
($\log [N_{\rm H}/{\rm cm^{-2}}] < 22$ [{\it open red circles\/}], $\log [N_{\rm
H}/{\rm cm^{-2}}] =$~22--23 [{\it open green triangles\/}], $\log [N_{\rm
H}/{\rm cm^{-2}}] > 23$ [{\it open blue squares\/}], and no measured $N_{\rm
H}$ due to lack of redshift measurements [{\it filled gray circles\/}]).  As in
Figure~9, AGN number-count predictions from G07 have been provided as dotted and
solid curves, corresponding respectively to the inclusion and exclusion of an
exponential decline in the AGN space density at $z > 2.7$ (model curve colors
correspond to symbol colors).
[{\it A machine-readable table of the $N_{\rm H}$-divided AGN cumulative number
counts data for all four bandpasses is provided in the electronic edition.}]
}
\end{figure*}

%
%
\begin{figure*}
\figurenum{12}
\centerline{
\includegraphics[width=16cm]{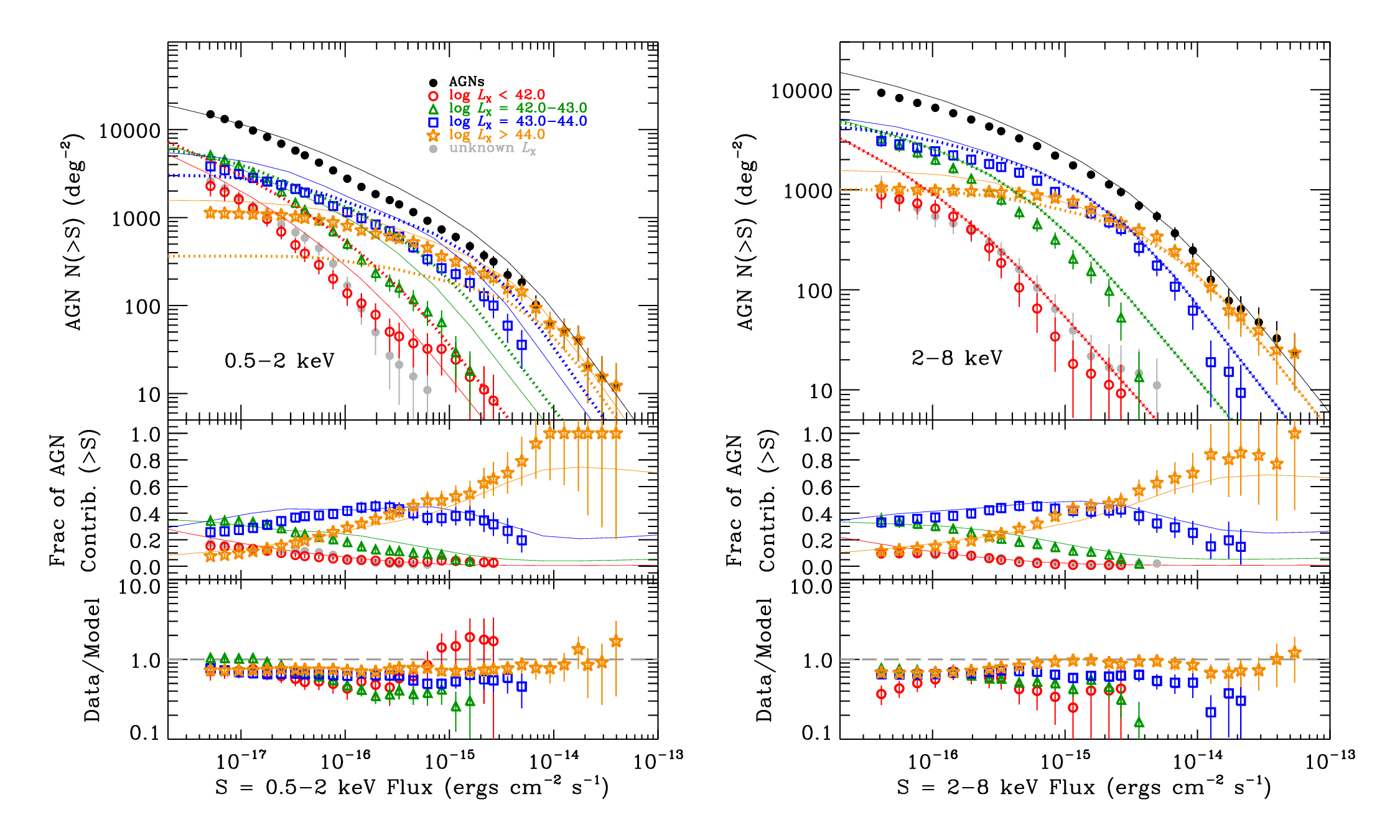}
}
\caption{
Same as Figure~9, but for AGNs divided into ranges of intrinsic
\hbox{0.5--8~keV} luminosity ($\log [L_{\rm X}/{\rm ergs~s^{-1}}] < 42$ [{\it
open red circles\/}], $\log [L_{\rm X}/{\rm ergs~s^{-1}}] =$~42--43 [{\it open
green triangles\/}], $\log [L_{\rm X}/{\rm ergs~s^{-1}}] =$~43--44 [{\it open
blue squares\/}], $\log [L_{\rm X}/{\rm ergs~s^{-1}}] > 44$ [{\it open orange
stars\/}], and no measured $L_{\rm X}$ due to lack of redshift measurements
[{\it filled gray circles\/}]).  As in Figure~9, AGN number-count predictions
from G07 have been provided as dotted and solid curves, corresponding
respectively to the inclusion and exclusion of an exponential decline in the
AGN space density at $z > 2.7$ (model curve colors correspond to symbol
colors).
[{\it A machine-readable table of the $L_{\rm X}$-divided AGN cumulative number
counts data for all four bandpasses is provided in the electronic edition.}]
}
\end{figure*}

\subsection{AGN Number Counts by Redshift, Intrinsic Column Density, and X-ray Luminosity}

As discussed in $\S$3.1, AGNs are the majority of the \xray\ detected sources
and dominate the cumulative number counts over all fluxes.  The ultradeep
\chandra\ data and redshift information allow for first-order estimates of the
intrinsic absorption column densities, $N_{\rm H}$, and \hbox{0.5--8~keV}
luminosities, $L_{\rm X}$, of the detected AGNs.  Values of $N_{\rm H}$ and
$L_{\rm X}$ were estimated following the procedure outlined in $\S$3.4 of Xue
\etal\ (2010).  The observed \xray\ spectrum was modeled using {\ttfamily
xspec} (Arnaud~1996) and an assumed absorbed power-law model ({\ttfamily
zpow}~$\times$~{\ttfamily wabs}~$\times$~{\ttfamily zwabs}).  The Galactic
column density was set to $8.8 \times 10^{19}$~cm$^{-2}$ (see $\S$1), the
intrinsic photon index was fixed at $\Gamma_{\rm int} = 1.8$ (e.g., Tozzi
\etal\ 2006), and the best redshift estimate provided in the X11 catalog was
adopted.  Using this model, we determined the $N_{\rm H}$ value for each AGN
that reproduced the observed ratio of count rates between the HB and SB.  For
sources detected only in the FB an observed spectral index of $\Gamma_{\rm obs}
= 1.4$ was adopted.  This procedure will be effective for determining $N_{\rm
H}$ and $L_{\rm X}$ values for AGNs in the Compton-thin/relatively unobscured
regime (i.e., with $N_{\rm H} \simlt 10^{23}$--$10^{24}$~cm$^{-2}$), and is
less reliable for a minority of sources that are expected to be Compton thick.
In the Compton-thick regime, AGN spectra are significantly complex due to the
combination of absorption, reflection, scattering, and line emission features
that depend on metallicity, ionization state, and geometry (see, e.g., Murphy
\& Yaqoob~2009).  Since most of our sources are in the low-count regime (i.e.,
$\simlt$100 counts), such meaningful modeling is beyond our capabilities.
Using our power-law model, estimated $N_{\rm H}$, and intrinsic flux $f_{\rm X,
int}$ estimates, we computed $L_{\rm X} = 4\pi d_L^2 f_{\rm X, int}
(1+z)^{\Gamma_{\rm int}-2}$ for each AGN that had a measured redshift
(spectroscopic or photometric).  

In Figure~9, we display the breakdown of the AGN number counts in bins of
redshift for the SB ({\it left\/}) and HB ({\it right\/}), with the middle
panels showing the fractional contribution that sources in each redshift range
make to the total AGN number counts.  At the brightest fluxes, low-redshift ($z
< 1.5$) AGNs dominate; however, going to fainter fluxes, we observe increasing
contributions from \hbox{$z =$~1.5--3} AGNs.  At the flux limit of our survey,
AGNs at \hbox{$z =$~1.5--3} make up comparable contributions to the number
counts as the $z < 1.5$ AGNs (i.e., $\approx$40\%).  For AGNs at $z>3$, we find
a roughly steady fractional contribution to the number counts
($\approx$5--15\%) across the majority of the SB and HB flux ranges where they
are detected.  In Figure~10, we highlight the SB $z > 3$ AGN number counts and
compare them with recent studies from the literature.  We find that our number
counts measurements for $z>3$ AGNs are typically a factor of
\hbox{$\approx$1.5--2} higher than those found in the COSMOS survey fields
(i.e., Brusa \etal\ 2009; Civano \etal\ 2011) and in good agreement with those
from the early $\approx$4~Ms \cdfs\ study by Fiore \etal\ (2011).  The
difference between our values and those of the COSMOS studies is likely due to differences
in $z > 3$ AGN selection techniques, variations in multiwavelength depth and
photometric redshift completenesses, and field-to-field variance.  

In Figures~11 and 12, we show the breakdown of AGN number counts in bins of
intrinsic absorption column density, $N_{\rm H}$, and \hbox{0.5--8~keV}
luminosity, $L_{\rm X}$, for the SB and HB.  In both bandpasses, we find that
unobscured \xray\ luminous sources dominate the bright-end number counts.
Going to fainter fluxes, we find that more obscured and less \xray\ luminous
sources make larger contributions.  At the SB flux limit, AGNs with $N_{\rm H}
\simgt 10^{22}$~cm$^{-2}$ and $L_{\rm X} \simlt 10^{43}$~\lum\ make up the
majority contributions to the number counts.  At the HB flux limit, we
similarly find that AGNs with $N_{\rm H} \simgt 10^{22}$~cm$^{-2}$ dominate the number
counts; however, in terms of luminosity, AGNs with $L_{\rm X} \simgt
10^{43}$~\lum\ and $\simlt 10^{43}$~\lum\ make comparable contributions, with
the latter population increasing faster.

\subsection{Normal-Galaxy Counts by Redshift and Morphology}

As noted in $\S$~3.1, normal galaxies comprise a significant fraction
($\approx$32--46\%) of the source counts at the SB and FB flux limits.  \xray\ emission
from normal galaxies is expected to be produced by a variety of populations
that differ for different galaxy populations (see, e.g., Fabbiano~1989, 2006
for a review).  For late-type star-forming galaxies, populations associated
with a combination of young and old stellar populations are expected to provide
dominant contributions (i.e., high-mass and low-mass \xray\ binaries,
supernovae and their remnants, hot gas from starburst flows, and young stars;
Colbert \etal\ 2004; Iwasawa \etal\ 2009, 2011; Lehmer \etal\ 2010;
Pereira-Santaella \etal\ 2011; Mineo \etal\ 2012).  For passive early-type
galaxies, \xray\ emission is dominated by low-mass \xray\ binaries (LMXBs) and
hot \xray\ emitting gas (e.g., O'Sullivan \etal\ 2001; Gilfanov~2004; Boroson
\etal\ 2011).  In this section, we study the normal-galaxy number counts as a
function of redshift and morphology.  

We measured the number counts of normal galaxies in two redshift bins divided
at the median redshift \hbox{$z_{\rm median} \approx 0.6$}.  In Figure~13, we
show the HB and SB number counts for normal galaxies at $z \simlt 0.6$ and $z
\simgt$~0.6.  In the SB, we find that low-redshift ($z \simlt 0.6$) normal
galaxies dominate the bright-end counts; however, below flux levels of
\hbox{$\approx$(1--2)}~$\times 10^{-17}$~\flux, high-redshift ($z \simgt 0.6$)
normal galaxies become the dominant galaxy population.  By contrast,
normal-galaxy number counts in the HB are dominated by low-redshift normal
galaxies across the full flux range, with $\simlt$20\% of the number counts
being from the high-redshift population.

To characterize the number counts of normal galaxies divided by morphology, we
made use of the multiwavelength data to classify the normal galaxies broadly as
either late-type or early-type galaxies.  We matched the 169 \xray\ detected
normal galaxies to the optical source catalogs constructed by Xue \etal\
(2010), which provide rest-frame optical magnitudes based on SED fitting.  To
distinguish between relatively blue late-type galaxies and red early-type
galaxies, we made use of the rest-frame $U-V$ color.  As described by Bell
\etal\ (2004), the $U$ and $V$ bandpass pair straddle the 4000~\AA\ break and
as a color provide a first-order indicator of the stellar age of the galactic
stellar population.  We applied the empirically calibrated redshift-dependent
color division calculated in $\S$~5 of Bell \etal\ (2004) to separate active
blue and passive red galaxy populations.  This division was set at
\begin{equation}
(U-V)_{\rm rest} = 1.15 - 0.31z - 0.08(M_V+20.7),
\end{equation}
such that galaxies redward and blueward of this division are candidate
early-type and late-type galaxies, respectively.  We note that a single color
division will not perfectly isolate the early-type and late-type galaxy
populations.  For example, edge-on or dusty late-type galaxies may be
artificially reddened and therefore classified as being early-type galaxies
(see, e.g., Cardamone \etal\ 2010; Lusso \etal\ 2011).  Furthermore, intrinsic
scatter in the optical color distributions near the color cut will sometimes
displace galaxies to the wrong side of the color division (e.g., ``green''
early-type galaxies that undergo low-levels of recent star-formation).  We
therefore visually inspected \hst\ $z_{850}$-band images of all 169 galaxies to
identify obvious cases where rest-frame colors provided incorrect morphological
classifications.  We found that 25 galaxies ($\approx$15\%) had obviously
incorrect initial morphological classifications that were corrected.

Out of the 169 normal galaxies, we classified 123 as late-type and 43 as
early-type galaxies.  In Figure~14, we show example \hst\ three-color
($B_{435}$, $V_{606}$, and $z_{850}$) images of 10 early-type ({\it
top\/}) and 10 late-type ({\it bottom\/}) galaxies that were in the normal
galaxy sample.  Following the prescriptions discussed in $\S$~2, we computed
number counts for each of these populations.  In Figure~15, we show the number
counts of the normal galaxies broken down by optical morphology.  
The normal-galaxy number counts for early-type and late-type galaxies are
comparable at the brightest SB and HB flux levels.  Progressing to lower
fluxes, however, we find that the late-type galaxy population quickly rises and
dominates the normal-galaxy number counts, and at the survey flux limits,
late-type galaxies comprise $\approx$80\% and $\approx$90\% of the normal galaxy
number counts for the SB and HB, respectively (see lower panels of Fig.~15).

\subsection{Contributions to the Cosmic X-ray Background}

With the number-count estimates derived above, we can measure the corresponding
contributions each source type and subtype make to the extragalactic CXRB.
Hereafter, we adopt CXRB intensities $\Omega_{\rm CXRB}$ of $(8.15 \pm 0.58)
\times 10^{-12}$~ergs~cm$^{-2}$~s$^{-1}$~deg$^{-2}$ and $(1.73 \pm 0.23) \times
10^{-11}$~ergs~cm$^{-2}$~s$^{-1}$~deg$^{-2}$ for the SB and HB, respectively.
These CXRB intensities were computed by summing components from (1) the
$\approx$1~Ms \cdfs\ unresolved background intensity ($\Omega_{\rm unres}$;
Hickox \& Markevitch~2006), (2) the intensity of faint $\approx$1~Ms \cdfs\
sources ($\Omega_{\rm faint}$; below $\approx$$5 \times 10^{-15}$~\flux\ and
$\approx$$1.4 \times 10^{-14}$~\flux\ for the SB and HB, resepectively) as
derived by Hickox \& Markevitch~(2006; see their Table~5), and (3) the
bright-source intensity ($\Omega_{\rm bright}$) derived from the best-fit
number-counts relations by Kim \etal\ (2007), which are based on $\sim$5500
\xray\ sources from the \chandra\ Multiwavelength Project (ChaMP; e.g., Green
\etal\ 2004; Kim \etal\ 2004a,b).  In the bright-source case, $\Omega_{\rm
bright} = \int_{S_{\rm faint}}^{S_{\rm bright}} S^\prime (dN/dS^\prime)
dS^\prime$, where $S_{\rm faint}$ is the brightest flux used to compute the
faint-end counts, $S_{\rm bright} = 10^{-11}$~\flux\ (for both the SB and HB),
and $dN/dS^\prime$ is the best-fit differential number counts model from
Table~3 of Kim \etal\ (2007).  Thus, we derived $\Omega_{\rm CXRB} =
\Omega_{\rm unres} + \Omega_{\rm faint} + \Omega_{\rm bright}$ for the SB and
HB.  For the FB and UHB, we adopted values of $(2.54 \pm 0.24) \times
10^{-11}$~\intensity\ (the sum of the SB and HB CXRB intensities) and $(1.04
\pm 0.23) \times 10^{-11}$~\intensity, respectively.  The UHB CXRB intensity
was computed by converting the HB intensity to UHB assuming a power-law SED
with $\Gamma = 1.4$ (see Moretti \etal\ 2009).

In Table~2 (col.~11--14), we provide the intensities and fractional CXRB
contributions that sources in various subcategories in the $\approx$4~Ms \cdfs\
provide.  We find that the vast majority of the CXRB in each band is expected
to be produced by AGNs, with normal galaxies producing only a small fraction
(\hbox{$\approx$0.5--3.0\%}) of the intensities.  These results are similar to
those found by B04, although revised to fainter flux levels.

Due to the relatively small solid-angle coverage of the \cdfs\ survey
($\approx$465~arcmin$^2$), we are unable to characterize the number counts for
the relatively rare \xray\ bright source population (i.e., $S
\simgt$~[1--10]~$\times 10^{-14}$~\flux).  To characterize the bright-end
counts for each source type and subtype listed in Table~2 (col.~2) would
require a wider survey with equivalent multiwavelength data to that used here
for the $\approx$4~Ms \cdfs.  Despite this limitation, we can utilize our
number-counts estimates and the total bright-end CXRB intensities (i.e.,
$\Omega_{\rm bright}$) to estimate the total resolved CXRB intensities.  In the
first row of Table~2, we provide the bright-end corrected resolved source
intensities and fractional contributions to the CXRB.  We find that
$\approx$76--88\% of the CXRB intensity can be attributed to \xray\ point
sources, with the resolved fraction appears to increase with median bandpass
energy; however, a constant or decreasing resolved fraction with energy is not
formally ruled out.  Comparisons with past studies indicate a variety of levels
of consistency with this result.  For example, the studies of Hickox \&
Markevitch~(2006) and Kim \etal\ (2007), which were used to estimate the total
CXRB intensities and bright-end corrections for the \cdfs, find similar
resolved CXRB fractions (using \hbox{$\approx$1--2~Ms} depth CDF data) over the same
energy ranges studied here.  However, the studies of Morretti \etal\ (2003),
B04, Worsley \etal\ (2005), and G08 find that the resolved CXRB fraction
appears to decline somewhat with energy, using data that reaches similar
\chandra\ depths (\hbox{$\approx$1--2~Ms}).  These studies adopt results from either
Morretti \etal\ (2003) and/or De~Luca \& Molendi~(2004) when estimating the
CXRB intensity and bright-end counts.  We note, however, that even in these
studies, the apparent decline in resolved CXRB with energy is less significant
in the CDF-S itself (see, e.g., Fig.~2 of Worsley \etal\ 2005).  We estimate
that between the \hbox{$\approx$1--2~Ms} and $\approx$4~Ms depths, the fraction
of the resolved CXRB has increased by only \hbox{$\approx$1--2\%}, which is much
smaller than the error bars of our measurements.  Therefore, differences between
previous resolved CXRB fractions are primarily related to either differing
assumptions about the CXRB intensity and bright-end counts or field-to-field
variations.  In support of the latter point, L08 find that the CDF-S number
counts are lower than those of the CDF-N by $\approx$25\% at the faintest
fluxes.  

\section{Comparisons with Models}

\subsection{The AGN Number Counts and the Evolution of Accreting SMBHs}

The redshift-dependent AGN number counts (Figs.~9 and 10) provide a directly
observable signature of the evolution of AGN activity in the Universe.  Since
AGNs are the brightest \xray\ detected population, many previous investigations
have focused on measuring their cosmic evolution (see Brandt \& Hasinger 2005
for a review), and in particular, the evolution of their \xray\ luminosity
function (XLF; e.g., Cowie \etal\ 2003; Ueda \etal\ 2003; Hasinger \etal\ 2005;
Silverman \etal\ 2008; Ebrero \etal\ 2009; Aird \etal\ 2010).  For example,
Hasinger \etal\ (2005) made use of a variety of \rosat, \xmm, and \chandra\
surveys with highly-complete optical/near-IR redshift measurements to compute
the observed evolution of the AGN XLF.  This investigation found that \xray\
selected AGNs appear to undergo luminosity-dependent density evolution in their
populations, where the space density of low-luminosity (Seyfert-type) AGNs
peaks at $z \simlt 1$ and the highly luminous (QSO-type) AGN space density
peaks at $z \approx 2$.  These direct measurements of the observed XLF
evolution of AGN activity in the Universe provide a useful constraint on the
SMBH accretion history; however, the observed \xray\ bandpass used to make
these measurements (i.e., \hbox{$\approx$0.5--2~keV}) is susceptible to
extinction and therefore populations of AGNs with intrinsic \xray\ absorption
columns of $N_{\rm H} \simgt 10^{22}$~cm$^{-2}$ will have under-represented
intrinsic accretion activity.

%
%
\begin{figure*}
\figurenum{13}
\centerline{
\includegraphics[width=16cm]{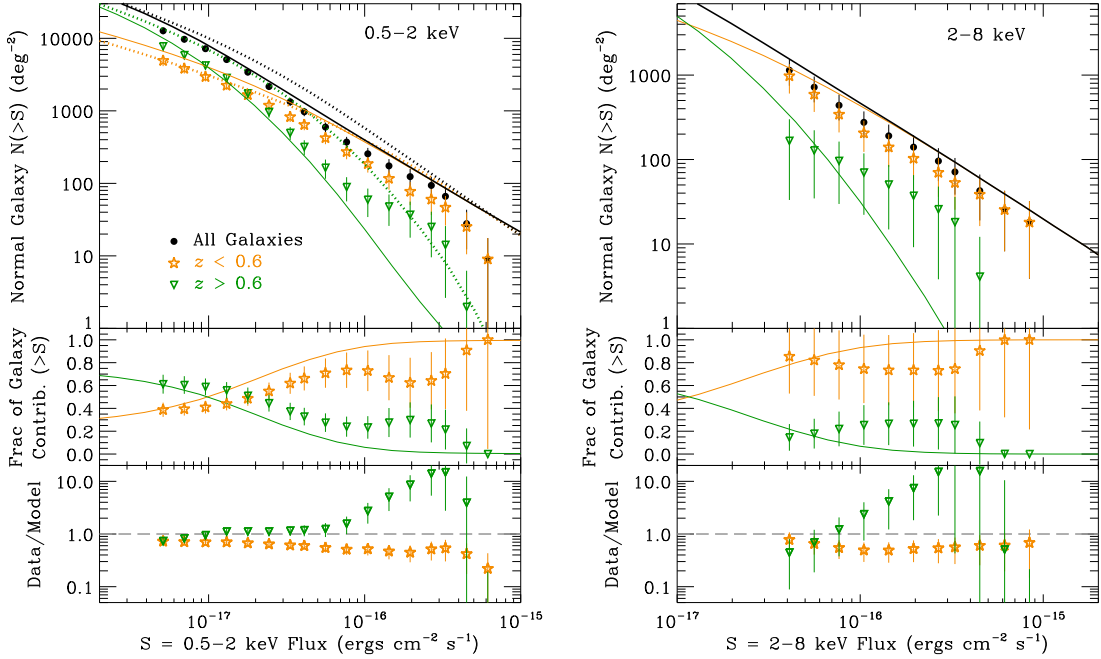}
}
\caption{
{\it Top Panels}: Cumulative normal-galaxy number counts ({\it filled black
circles\/}) in the SB ({\it left\/}) and HB ({\it right\/}) broken down by
contributions from galaxies at redshifts $z \simlt 0.6$ ({\it open orange
stars\/}) and $z \simgt$~0.6 ({\it open green upside-down triangles\/}).
The redshift division between the two normal-galaxy samples was made at the
median galaxy redshift.  The dotted curves in the left panel show the predicted
$z \simlt 0.6$ ({\it orange dotted curve\/}) and $z \simgt$~0.6 ({\it green
dotted curve\/}) number counts based on the observed XLF pure-luminosity
evolution parameterization derived by Ptak \etal\ (2007) using observed XLFs
out to $z \approx 0.75$.  At $z \simgt 0.6$, these predictions are based on an
extrapolation of the XLF parameterization out to $z \approx 2$.   The solid
curves in the left and right panels show the predicted number counts for $z \simlt 0.6$ ({\it
orange solid curves\/}) and $z \simgt$~0.6 ({\it green solid curves\/}) galaxy
populations based on the observed evolution of stellar mass functions converted
to XLFs using \xray\ scaling relations (see $\S$4.2 for details).
{\it Middle Panels}: The fractional contributions that $z \simlt 0.6$ and $z
\simgt$~0.6 galaxy populations make to the total galaxy number counts.  
{\it Bottom Panels}: The ratio between the data and predictions based on the
solid curves in the top panels.
[{\it A machine-readable table of the redshift-divided normal galaxy cumulative
number counts data for all four bandpasses is provided in the electronic
edition.}]
}
\end{figure*}

%
%
\begin{figure*}
\figurenum{14}
\centerline{
\includegraphics[width=18cm]{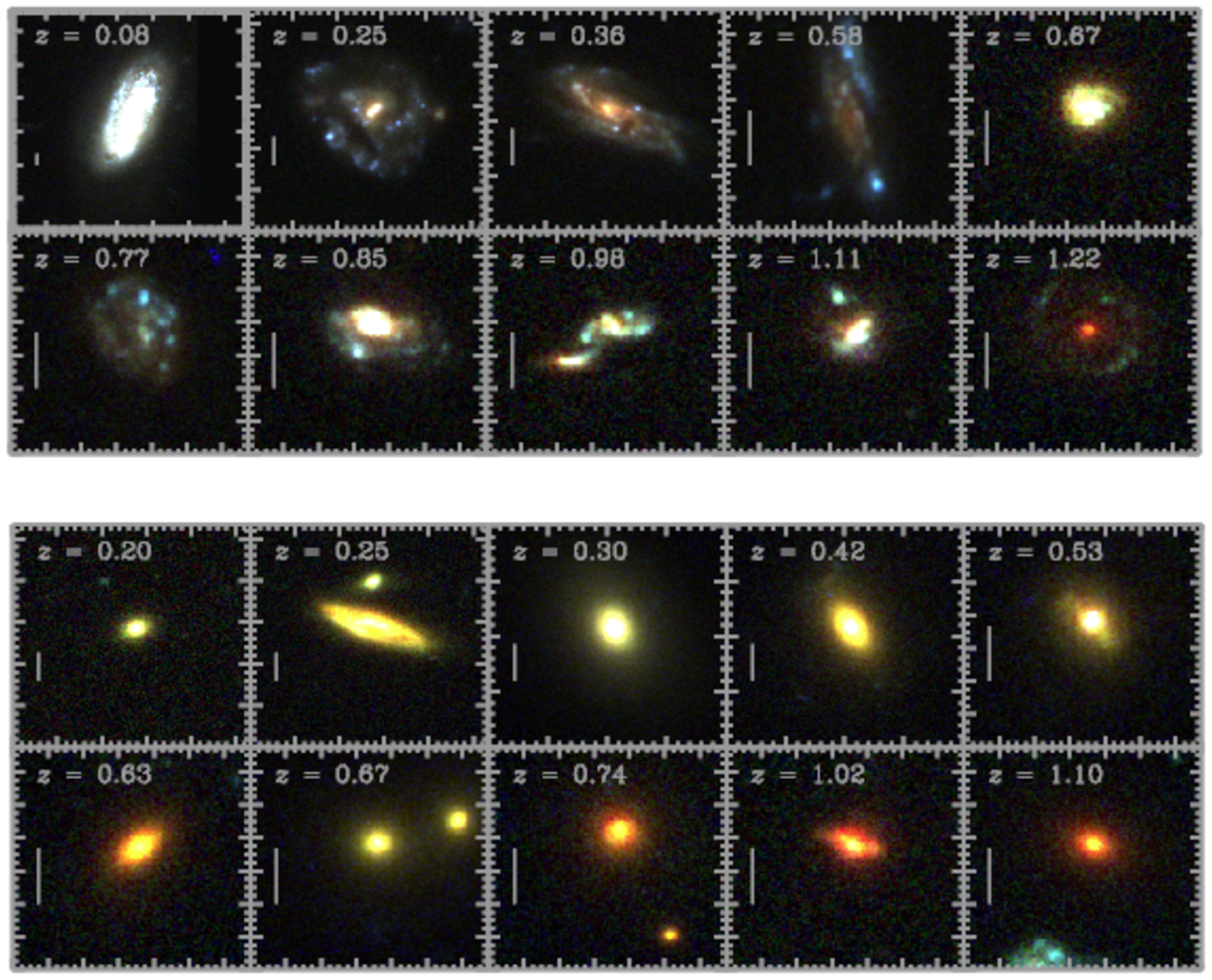}
}
\caption{
Three-color ($B_{435}$, $V_{606}$, and $z_{850}$) \hst\ images of a sample of
\xray\ detected normal galaxies ordered by redshift.  The top panels show 10
example late-type galaxies and the bottom panels show 10 example early-type
galaxies.  The angular extent of each image has been chosen to subtend a
physical distance of $\approx$30~kpc on a side in the rest-frame of each
galaxy.  The redshift of each source is indicated in the upper left and a
1~arcsec bar has been placed in the lower-left hand corner to show angular
scale.  The images are centered on the coordinates corresponding to each
galaxy, which were taken from columns~18 and 19 of Table~3 in X11.  These
positions were obtained primarily from VLT WFI observations, and therefore
small ($\simlt$0.1--0.2~arcsec) offsets are sometimes seen in these \hst\
images (note the angular scale). 
}
\end{figure*}

Using the unresolved hard \xray\ ($\approx$3--100~keV) background intensity
spectrum as a constraint on the obscured ($N_{\rm H} \approx
10^{22}$--$10^{24}$~cm$^{-2}$) and Compton-thick ($N_{\rm H} \simgt
10^{24}$~cm$^{-2}$) AGN populations, Gilli \etal\ (2007; hereafter, G07) used
the observed \xray\ luminosity function evolution from Hasinger \etal\ (2005),
local distributions of the luminosity-dependent obscured-to-unobscured AGN
fraction, and a model distribution of AGN population column densities to form
phenomenological models describing the intrinsic evolution of the AGN \xray\
luminosity function (see also Ueda \etal\ 2003; Draper \& Ballantyne~2009;
Treister \etal\ 2009 for additional prescriptions).  These models make direct
predictions for the number count distributions of sources with different ranges
of $z$, $N_{\rm H}$, and intrinsic \hbox{0.5--8~keV} luminosity $L_{\rm
X}$,\footnote{The G07 phenomenological model number-counts predictions can be
found at http://www.bo.astro.it/$\sim$gilli/counts.html.} and can be directly
compared with our observed AGN number-count subpopulations (see Figs.~9--12).

In \hbox{Figures~9--12}, we have highlighted the number-count predictions for
$z$, $N_{\rm H}$, and $L_{\rm X}$ selected AGNs by G07 appropriate for the
Hasinger \etal\ (2005) XLF as observed ({\it solid curves\/}) and after
applying a high-redshift ($z > 2.7$) exponential decline to the XLF ({\it
dotted curves\/}; see Schmidt \etal\ 1995 for motivation).  In general, we
found better agreement by not including the exponential declining term
(i.e., the {\it solid curves} in Figs.~9--12).  Although, since the majority of
the AGNs in this study are at $z \simlt 2.7$, there is generally little
difference between the models that include and do not include an exponentially
declining term.  For clarity in comparisons, in Figs.~9, 11, and 12 we have
provided lower panels showing the ratio between our observed number counts and
the G07 models without the exponential declining term.

For redshift-selected number counts, we found reasonable agreement (within a
factor of $\approx$2) between the observed number counts and the G07 model
across the entire SB and HB flux ranges (see Fig.~9).  We note that a
non-negligible fraction of the AGN population does not have redshifts
available.  Since the multiwavelength coverage over the entire \cdfs\ is
extensive and deep, sources without redshifts (neither spectroscopic nor
photometric) have very faint optical/near-IR counterparts, and are are
therefore good candidates for high-redshift ($z > 3$) AGNs.  If we add the
number counts from AGNs without redshifts to the $z > 3$ number counts, we find
improved agreement with the G07 model prediction (see Fig.~10 {\it
diamonds\/}).  For comparison, we have also plotted the predictions from the
recent work by Aird \etal\ (2010), which is based on \hbox{2--10~keV} XLF
measurements from the CDFs and the AEGIS-X surveys.  The Aird \etal\ (2010)
predictions are also consistent with our $z > 3$ AGN number counts; however, if
all of the AGNs with unknown redshifts are in the $z > 3$ range, our number
counts become more consistent with the G07 model.  Regardless of the redshifts
of the sources lacking redshift identifications, our number counts appear to be
a factor of $\simgt$2 higher than those predicted by the G07 model with
exponential decline.

For the number counts predictions based on $N_{\rm H}$, we find significant
discrepancies (factors of $\approx$1.5--3) between those observed and those
predicted by the G07 model (see bottom panels of Fig.~11).  This situation is
similar regardless of whether the exponential decline is applied or not.  We
suspect that some of the disagreement here will be due to (1) cosmic variance
of the populations, (2) our simplistic calculation of $N_{\rm H}$ based on band
ratios (see discussion in $\S$3.2), which differs from the spectral models from
G07 (see their Fig.~1), (3) the distribution of $N_{\rm H}$ values for sources
without redshifts, and (4) imperfections in the G07 model descriptions.
Regarding point~(3), we note that sources with unknown redshifts have a median
effective photon index of $\Gamma_{\rm median} \approx 0.7$, which is somewhat
harder (softer) than the $\Gamma_{\rm median} \approx 1.0$ ($\Gamma_{\rm
median} \approx 0.4$) for sources with $\log (N_{\rm H}$/cm$^{-2})$ = 22--23
($\log [N_{\rm H}$/cm$^{-2}] > 23$).  Therefore, we expect that most of the
sources with unknown redshifts will have $\log (N_{\rm H}$/cm$^{-2}) \simgt 23$
regardless of their redshifts.  This would improve the agreement between
observed and model number counts for $\log (N_{\rm H}$/cm$^{-2}) \simgt 23$.
Despite the apparent disagreements and limitations, we do find basic agreement
in the general trend: more heavily obscured AGNs have larger contributions at
fainter fluxes.

For number counts divided by intrinsic \hbox{0.5--8~keV} luminosity $L_{\rm
X}$, we find good (within a factor of $\approx$2) basic agreement between our
measurements and the G07 model (see Fig.~12); however, this is not true for the
decline model (see {\it dotted curves} in upper panels of Fig.~12). 

In summary, we find that the G07 phenomenological models provide a good
description of the redshift, column density, and intrinsic luminosity
distributions to the extent that we can constrain these values observationally.
Better measurements of the column densities and $L_{\rm X}$ would be needed to
improve these models further.  However, unless a significant fraction of
objects with photometric redshifts $z > 3$ prove to be lower redshift
contaminants, we do find that applying an exponential decline to the AGN XLFs
at $z > 2.7$, under the G07 parameterizations, provides a poor characterization
of the redshift and $L_{\rm X}$ divided number counts.

\subsection{The Rapidly Increasing Normal Galaxy Counts}

In $\S$~3.3, we presented number-counts estimates for the normal-galaxy
population and subpopulations separated by redshift and galaxy morphology.
Given that these \xray\ source populations are uniquely accessible to the
ultradeep survey regime probed by the \chandra\ Deep Fields, the evolution of
the normal-galaxy XLFs have not been as well characterized as those of AGN
populations (see $\S$~4.1).  However, initial investigations have placed
first-order constraints on the evolution of normal-galaxy XLFs out to $z
\approx 0.75$ (e.g., Norman \etal\ 2004; Georgakakis \etal\ 2007; Ptak \etal\
2007; Tzanavaris \& Georgantopolous~2008).  These investigations have
provided evidence for rapidly increasing late-type galaxy 0.5--2~keV XLF that
is consistent with pure luminosity evolution, where $L^* \propto (1 + z)^b$,
and \hbox{$b \approx$~1.5--3}.  By contrast, the early-type galaxy 0.5--2~keV
XLF appears to be changing more slowly with redshift, and in the Tzanavaris \&
Georgantopolous~(2008) study, the XLF is consistent with being constant or
decreasing with redshift (the value of \hbox{$b$} ranges from $\approx$~$-$2.3 to $+$2.6
in these studies).  However, these results are based on shallower data
than the $\approx$4~Ms \cdfs, and thus far provide only estimates for how the
bright-end of the galaxy XLFs evolved out to $z \approx 0.75$.  In the
$\approx$4~Ms \cdfs, $\approx$29\% of the normal galaxies are at $z \simgt
0.75$, and the majority of these sources are only accessible by this survey.
In the paragraphs below, we interpret the number counts of the normal galaxy
subpopulations (e.g., see Figs.~13 and 15) in the context of what we expect
from previous estimates of the galaxy XLF evolution and the evolution of galaxy
properties (e.g., SFR and stellar mass) and \xray\ scaling relations (e.g., the
\xray/SFR relation).

Following the formalism of Ranalli \etal\ (2005), the normal-galaxy number
counts can be expressed as the following double integral:
\begin{equation}
N(>S) = K_{\rm sr}^{\rm deg} \int_{z_{\rm min}}^{z_{\rm max}} dz \int_{L_{\rm
min}(S)}^{L_{\rm max}} \varphi_{\rm X}(\log L_{\rm X}, z) \; d\log L_{\rm X}
\frac{dV}{dz d\Omega},
\end{equation}
where $K_{\rm sr}^{\rm deg} = 3.05 \times 10^{-4}$ converts from sr$^{-1}$ to
deg$^{-2}$, $\varphi_{\rm X}$ is the redshift-dependent XLF, $\frac{dV}{dz
d\Omega}$ is the cosmology-dependent differential volume element (comoving
volume per unit redshift per unit solid angle), and [$z_{\rm min}$, $z_{\rm
max}$] and [$L_{\rm min}(S)$, $L_{\rm max}$] are respectively redshift and
luminosity integration limits.  The value of $L_{\rm min}(S)$ is the rest-frame
luminosity corresponding to a flux $S$ and redshift $z$.  In the upper-left
panels of Figures~13 and 15, the dotted curves show \hbox{0.5--2~keV} number
count predictions based on previous measurements of the evolution of galaxy
XLFs for redshift and morphology divided galaxy populations, respectively.
These curves were computed using the Ptak \etal\ (2007) redshift-dependent
Schechter parameterizations of the observed galaxy XLFs, which assume
pure-luminosity evolution of the form $L^* \propto (1+z)^{2.3}$ and
$(1+z)^{1.6}$ for late-types and early-types, respectively.  We also
experimented with using a log-normal parameterization of the galaxy XLF and
found similar results.  When estimating galaxy number counts predictions, we
used equation~9 and extrapolated the Ptak \etal\ (2007) parameterizations out
to $z \approx 2$.  The Ptak \etal\ (2007) XLFs were directly measured for
galaxies out to $z \approx 0.75$ (using galaxies with redshifts as large as $z
\approx 1.2$) and are in good agreement with other galaxy XLF measurements from
the literature (e.g., Norman \etal\ 2004; Tzanavaris \& Georgantopolous~2008).
The uncertainties on the normalization of the XLF parameterization is a factor
of $\approx$2.  We find that, within the uncertainties, the XLFs provide
reasonable predictions for the $z \simlt 0.6$ (Fig.~13, {\it orange dotted
curve and stars\/}) and late-type galaxy (Fig.~15, {\it blue dotted curve and
squares\/}) number counts.  For these two samples, the largest differences come
from the $z \simlt 0.6$ sample at 0.5--2~keV fluxes $\simgt$$3 \times
10^{-17}$~\flux, which may be due to differences in galaxy classification of
\xray\ detected sources, cosmic variance, and statistical errors associated
with the XLF fits.  We find that the extrapolated XLFs appear to overpredict
the $z \simgt 0.6$ (Fig.~13, {\it green dotted curve and upside-down
triangles\/}) and early-type galaxy (Fig.~15, {\it red dotted curves and
triangles\/}) number counts by more than a factor of two, suggesting that our
extrapolation of the early-type galaxy XLF evolution out to $z \approx 2$ is an
overestimate.  Therefore, it is possible that the {\itshape rate} by
which the early-type galaxy XLF rises with redshift decreases somewhat going
from $z \approx$~0.75--2 and/or the adopted rate of increase is too high
(e.g., see Tzanavaris \& Georgantopolous~2008 who find a lower rate of
increase).  To obtain independent predictions of the galaxy XLFs and number
counts for galaxies out to $z \approx 2$, we can use the combination of \xray\
scaling relations (e.g., the \xray/SFR relation) and the redshift evolution of
galaxy physical properties (e.g., SFR and stellar mass) that are measured using
multiwavelength observations.

%
%
\begin{figure*}
\figurenum{15}
\centerline{
\includegraphics[width=16cm]{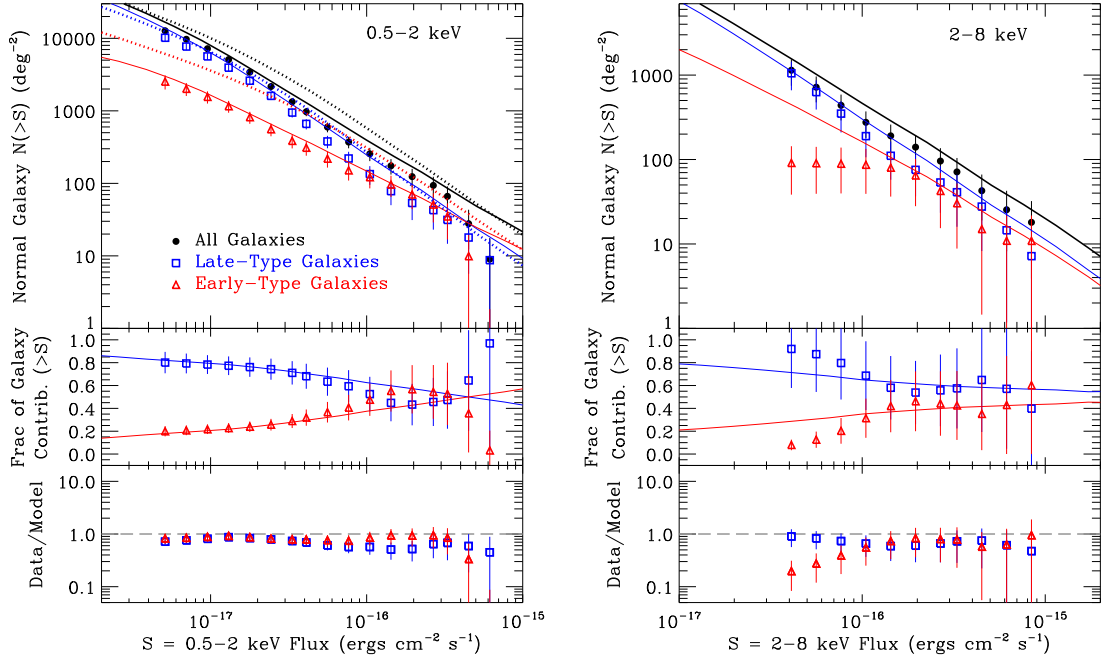}
}
\caption{
{\it Top Panels}: Similar to Figure~13, but for normal galaxies divided into late-type ({\it open
blue squares\/}) and early-type ({\it open red triangles\/}) galaxy
populations.  The classification of galaxy morphology is explained in $\S$3.3.
The dotted curves in the upper left panel show the predicted late-type
({\it blue dotted curve\/}) and early-type ({\it red dotted curve\/}) galaxy
number counts based on the observed XLF pure luminosity evolution
parameterization derived by Ptak \etal\ (2007) using observed XLFs out to $z
\approx 0.75$.  These predictions are based on extrapolation of the XLF
parameterization out to $z \approx 2$.  The solid curves in the left and
right panels show the predicted number counts for late-type ({\it blue
solid curve\/}) and early-type ({\it red solid curve\/}) galaxy populations
based on the observed mass functions converted to XLFs using \xray\ scaling
relations (see $\S$4.2 for details).
{\it Middle Panels}: The fractional contributions that late-type and early-type
galaxies make to the total galaxy population.
{\it Bottom Panels}: The ratio between the data and predictions based on the
solid curves in the top panels.  The reasonable agreement between the
number-counts data and models, both here and in Figure~13, indicates that the
rapidly rising galaxy counts are largely attributed to the evolution of galaxy
physical properties (e.g., stellar mass and star-formation rate) with little
redshift evolution in scaling relations between these physical properties and
\xray\ power output.
[{\it A machine-readable table of the morphology-divided normal galaxy
cumulative number counts data for all four bandpasses is provided in the
electronic edition.}]
}
\end{figure*}

Recent \xray\ stacking investigations of large normal-galaxy populations
selected by morphology and galaxy physical properties (e.g., SFR and stellar
mass) have now provided insight into how the \xray\ emission from normal galaxies
evolves relative to galaxy physical properties out to $z \approx$~1--4.
Lehmer \etal\ (2008) reported that late-type star-forming galaxies, the most
numerous galaxy population in the Universe, have mean $L_{\rm X}$/SFR values
that are roughly constant out to $z \approx 1.4$.  Further \xray\ stacking of
distant $z \approx$~1.5--4 Lyman break galaxies appears to show similar mean
$L_{\rm X}$/SFR values (e.g., Brandt \etal\ 2001; Lehmer \etal\ 2005; Laird
\etal\ 2006; Cowie \etal\ 2011; Basu-Zych \etal\ 2012, in-prep).  Similarly,
Vattakunnel \etal\ (2012) have studied $z \simlt 1.2$ normal galaxies in the
$\approx$4~Ms \cdfs\ at \xray\ and radio wavelengths, and found that the
$L_{\rm X}$--$L_{\rm 1.4~GHz}$ correlation for these galaxies is consistent
with that expected from the local $L_{\rm X}$/SFR relation, albeit with
significant scatter ($\approx$0.4~dex).  Additional indirect support for a
roughly constant $L_{\rm X}$/SFR ratio with redshift has been provided by
Dijkstra \etal\ (2011) who used the observed SFR density evolution of the
Universe to show that the unresolved SB CXRB can be fully explained by a
$L_{\rm X}$/SFR ratio $\propto (1+z)^b$, where $b$ is constrained to be less
than 1.4.  

For early-type galaxies, Lehmer \etal\ (2007) and Danielson \etal\ (2012)
showed that the soft-band \xray\ luminosity per unit $B$-band luminosity
($L_{\rm X}/L_B$) appears to undergo only mild evolution ($\propto
[1+z]^{1.2}$) to $z \approx 1.2$ for the most optically luminous galaxies,
which is expected to be due to \xray\ emitting hot gas.  In contrast,
however, lower-luminosity early-type galaxies have been found to have rising
mean $L_{\rm X}/L_B$ values with increasing redshift, which is expected to be
due to a fading LMXB population toward the present day (Lehmer \etal\ 2007;
Hornschemeier \etal\ 2012, in prep).  Similar expectations have been found from
\xray\ binary population-synthesis models, which predict an evolution of
$L_{\rm X}/L_B \propto (1+z)^2$ at $z \simlt 1$ (Fragos \etal\ 2012, in prep;
see also Ghosh \& White 2001).

The above investigations provide estimates for how scaling relations between
\xray\ luminosity and galaxy physical properties have evolved out to $z
\approx$~1--4, where the majority of the $\approx$4~Ms \cdfs\ normal galaxies
are detected.  It seems that only the scaling relation between LMXB emission
and stellar mass undergoes significant evolution with cosmic time; however,
these populations are expected to dominate only in early-type galaxies, which
make up a minority of the \xray\ detected sources in the $\approx$4~Ms \cdfs.
It is therefore likely that the evolving normal-galaxy XLF and rising galaxy
number counts, observed here and in previous studies, is largely due to the
rapid evolution of galaxy physical properties.  Recent multiwavelength surveys
(e.g., COSMOS; Scoville \etal\ 2007) have provided significant new constraints
on the evolution of optical/near-IR luminosity, SFR, and stellar-mass
functions.  These measurements can be combined with scaling relations between
\xray\ luminosity and galaxy physical properties to make predictions for the
evolution of the normal-galaxy XLFs and number counts for galaxies out to
$z \approx 2$.

Following Avni \&
Tananbaum (1986), $\varphi_{\rm X}$ can be estimated through the following
transformation:
\begin{equation}
\varphi_{\rm X}(\log L_{\rm X}, z) = \int_{-\infty}^{\infty}
\varphi_\vartheta(\log \vartheta, z) P(\log L_{\rm X} \vert \log \vartheta) \;
d\log \vartheta,
\end{equation}
where $\vartheta$ represents a galaxy physical property (e.g., optical
luminosity, stellar mass, or SFR) by which $\varphi_\vartheta(\log \vartheta,
z)$ has been measured, and $P(\log L_{\rm X} \vert \log \vartheta)$ is the
probability distribution for observing $L_{\rm X}$ given the value $\vartheta$.

We make use of the redshift-dependent stellar-mass functions measured by Ilbert
\etal\ (2010) for late-type and early-type galaxies out to $z \approx 2$ (i.e.,
$\vartheta = M_\star$).  To convert $\varphi_{M_\star}(\log M_\star, z)$ to
$\varphi_{\rm X}(\log L_{\rm X}, z)$ via equation~9 requires knowledge of the
transformation between \xray\ luminosity and stellar mass (i.e., $P[\log L_{\rm
X} \vert \log M_\star]$).  This function is expected to depend on galaxy
morphology.  In its most general form, we can express the transformation
function as follows:
\begin{equation}
P(\log L_{\rm X} \vert \log M_\star) = \frac{1}{\sqrt{2\pi} \sigma} \exp \left[
{-\frac{(\log L_{\rm X, mod} (M_\star) - \log L_{\rm X})^2}{2 \sigma^2}}
\right],
\end{equation} 
where $L_{\rm X, mod}(M_\star)$ is the predicted \xray\ luminosity given a
value of the stellar mass $M_\star$ and $\sigma$ represents the scatter in the
relation.  For normal late-type galaxies in the local Universe, the \xray\
luminosity has been shown to correlate strongly with SFR (e.g., Persic \&
Rephaeli~2007; Lehmer \etal\ 2010; Pereira-Santaella \etal\
2011; Symeonidis \etal\ 2011; Mineo \etal\ 2012), and as discussed above, this correlation is
observed to hold out to $z \simgt 1$.  We made use of the Lehmer \etal\ (2010)
\xray/SFR relation for late-type star-forming galaxies in the local Universe:
\begin{equation}
L_{\rm 2-10~keV}  =  \alpha M_\star + \beta {\rm SFR}
\end{equation}
where $\alpha = 9.08 \times 10^{28}$~ergs~s$^{-1}$~$M_\odot^{-1}$ and $\beta =
1.62 \times 10^{39}$~ergs~s$^{-1}$~$(M_\odot {\rm yr}^{-1})^{-1}$, $M_\star$ is
in units of $M_\odot$, SFR is in units of $M_\odot$~yr$^{-1}$, and $L_{\rm
2-10~keV}$ is in units of \lum.  Studies of local and distant galaxy
populations have shown that the SFR for late-type star-forming galaxies is
further strongly correlated with $M_\star$ out to $z \approx 2$ (e.g., Daddi
\etal\ 2007; Elbaz \etal\ 2007; Peng \etal\ 2010), albeit with a rapidly
evolving normalization (i.e., mean SFR/$M_\star$) with redshift (e.g., Zheng
\etal\ 2007; Karim \etal\ 2011; Cen~2011).  To convert SFR to $M_\star$ for
late-type star-forming galaxies, we adopted the relation provided by Karim
\etal\ (2011): 
\begin{equation}
{\rm SFR}/M_\star  = c_0 (1 + z)^{3.5} (M_\star/10^{11} M_\odot)^{\beta_{\rm
SFR}}
\end{equation}
where $c_0 \approx 0.0263$~Gyr$^{-1}$ and $\beta_{\rm SFR} \approx -0.4$ are
fitting constants.  Combining this relation with equation~12, we arrive at the
following expression:
\begin{equation}
L_{\rm 2-10~keV} = \alpha M_\star + \gamma (1+z)^{3.5} M_\star^{0.6}
\end{equation}
where $\gamma = 1.07 \times 10^{33}$, $M_\star$ is in units of $M_\odot$, and
$L_{\rm 2-10~keV}$ is in units of \lum.  When computing number-count
predictions, we converted the \hbox{2--10~keV} luminosities provided here to
other bandpasses assuming a power-law SED with $\Gamma = 1.9$, the average
stacked SED of the \xray\ detected normal galaxies (Young \etal\ 2012).  The
overall scatter (accounting for both SFR/$M_\star$ and $L_{\rm X}$/SFR
relations) in this relation is estimated to be $\sigma \approx 0.4$~dex.

For normal early-type galaxies, the \xray\ luminosity has been shown to
correlate with $K$-band luminosity, $L_K$, which provides a direct proxy for
galaxy stellar mass (e.g., Gilfanov~2004; Boroson \etal\ 2011).  We converted
the Boroson \etal\ (2011) $L_{\rm X}$/$L_K$ relations for early-type galaxies
to $L_{\rm X}$/$M_\star$ relations assuming a single mass-to-light ratio
\hbox{$\log M_\star/L_K \approx -0.2$} (where $M_\star$ and $L_K$ are in solar
units) characteristic of early-type galaxies.  These relations are separated
into contributions from LMXBs and hot \xray\ emitting gas:
\begin{eqnarray}
\log L_{\rm 0.3-8~keV} ({\rm LMXB}) \approx 29.2 + \log M_\star + 2 \log(1+z)
\nonumber \\
\log L_{\rm 0.3-8~keV} ({\rm gas}) \approx -6.07  + 4.03 \log M_\star,
\end{eqnarray}
where the $2 \log(1+z)$ term accounts for the expected redshift evolution for
LMXBs in early-type galaxies (see discussion above).  When computing
number-counts predictions, we converted the \hbox{0.3--8~keV} relations to
other bandpasses assuming a power-law SED with $\Gamma = 1.8$ for LMXBs and an
{\ttfamily apec} plasma model with an $L_{\rm 0.3-8~keV}$ dependent temperature
($T_{\rm X} \propto L_{\rm 0.3-8~keV}^{0.21}$) and solar abundances for the hot
gas.  The scatter in the LMXB and hot-gas relations are estimated to be
$\approx$0.2~dex and $\approx$1.0~dex, respectively (see Boroson \etal\ 2011).

The relations expressed in equations 14 and 15, combined with the stellar-mass
functions from Ilbert \etal\ (2010), were folded into equation~9 to obtain
redshift-dependent XLFs for late-type and early-type galaxies, respectively.
We measured the number counts of both populations in two redshift bins by
integrating equation~9 over the redshift intervals \hbox{$z \simlt 0.6$} and
\hbox{$z \simgt$~0.6} and summing the contributions from each galaxy morphology
type.  In Figures~13 and 15, we show our model number counts in bins of
redshift and morphology, respectively, and in the bottom panels, we plot the
ratio of our data to the model.  We remind the reader that the models
implemented here are not fits to the data in any way.  For the redshift-divided
normal galaxy samples, we find good agreement between our data and the models
at the SB and HB flux limits; however, for the number counts for normal
galaxies at $z \simgt$~0.6, we find significant differences at the bright-end
($f_{\rm 0.5-2~keV}$ and $f_{\rm 2-8~keV} \simgt 10^{-16}$~\flux).  The
bright-end number counts for $z\simgt$~0.6 galaxies are based on only
$\approx$1--5 objects; therefore, the differences with the models may be due to
either cosmic variance or misclassification of some fraction of these sources
(i.e., if some of these normal galaxies are actually AGNs).  Furthermore, the
general trends seen in the data and models (e.g., the increasing contributions
and faint-end dominance in the SB of $z \simgt$~0.6 galaxies) appear to be
broadly consistent with each other.  For the morphology-selected number counts,
we find very good agreement (less than a factor of $\approx$1.5 difference)
between the data and the model with the exception of the HB early-type galaxy
counts.  We note, however, that the HB number counts measurements for
early-type galaxies are based on small numbers of galaxies (four galaxies) and
are sensitive to the classifications discussed in $\S$3.1 above. 

In summary, we find that our models provide good characterizations of the
\xray\ number counts in redshift bins for both late-type and early-type galaxy
populations down to the flux limits of our survey (with the exception of $z \simgt
0.6$ normal galaxies with bright SB and HB fluxes and HB early-type galaxy
counts; however, see discussion above).  For late-type galaxies, this result
suggests that \xray\ scaling relations in the local universe appear to hold out
to at least $z \approx$~1--2, in agreement with previous investigations.  For
early-type galaxies, these results are consistent with there being little
redshift evolution in how hot gas \xray\ emission scales with stellar mass and
modest redshift evolution ($\propto [1+z]^2$) in the LMXB luminosity per
stellar mass.

\section{The Impending Dominance of Normal Galaxies: Predictions for Future
Deeper X-ray Observations}

%
%
\begin{figure}
\figurenum{16}
\centerline{
\includegraphics[width=9cm]{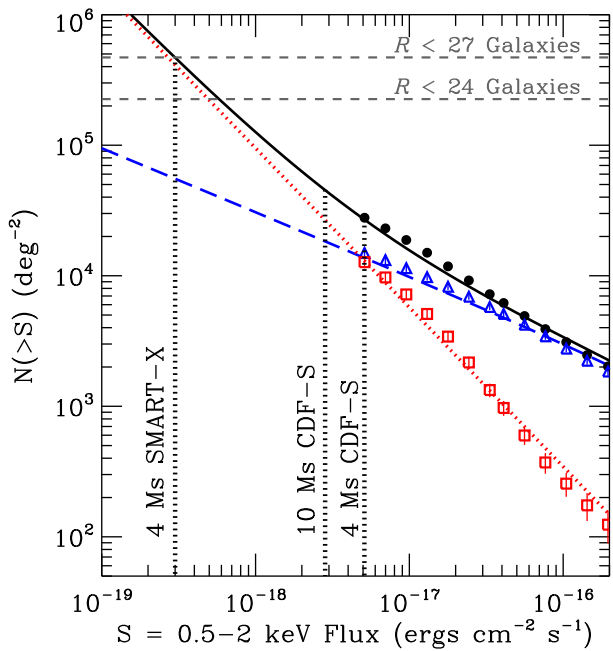}
}
\caption{
Measured SB number counts ({\it filled black circles\/}) with contributions
from AGNs ({\it open blue triangles\/}) and normal galaxies ({\it open red
squares\/}) shown.  Our best-fit parameterizations from equation~5 for all
sources, AGNs, and normal galaxies have been shown as solid black, dashed blue,
and dotted red curves, respectively.  An extrapolation of these models
indicates that the normal-galaxy number counts will overtake those of AGNs at
SB flux levels of $\approx$(3--5)~$\times 10^{-18}$~\flux.  The ultimate flux
limits of the $\approx$4~Ms \cdfs, a potential $\approx$10~Ms \cdfs, and future
$\approx$4~Ms {\it SMART-X} exposures have been indicated.  These
extrapolations indicate that a $\approx$10~Ms \cdfs\ would have SB number
counts at the flux limit that reach source densities of
$\approx$46,000~deg$^{-2}$ and are dominated by normal galaxies.  More
ambitiously, a $\approx$4~Ms {\it SMART-X} exposure would reach source
densities of $\approx$500,000~deg$^2$ and would have an $\approx$90\%
contribution from normal galaxies.  For reference, we have plotted the source
densities of galaxies with $R$-band magnitudes brighter than $R \approx 24$ and
$R \approx 27$ ({\it gray short-dashed lines\/}). 
}
\end{figure}

The number-counts results presented above show that normal galaxies play an
increasingly dominant role in the extragalactic \xray\ source population going
to fainter fluxes.  Our observations show that the cumulative SB and FB number
counts of normal galaxies are on the brink of surpassing those of AGNs (see
Fig.~5), and the slope of the galaxy number counts ($dN/dS$) already exceeds
that of AGNs (see Fig.~6).  At the SB flux limit of the $\approx$4~Ms \cdfs\,
normal galaxies are reaching the unprecedented source density of
$\approx$\galden~deg$^{-2}$, which constitutes $\approx$46\% of the total
number counts.  Given the comparable source densities of normal galaxies and
AGNs at the survey flux limit and the relatively sharp faint-end rise in the
galaxy number counts versus AGNs ($N[>S]$ power-law slope of $\beta^{\rm gal} -
1 \approx 1.2$ versus $\beta_1^{\rm AGN} -1 \approx 0.5$; see Table~1), it is
clear that going to fainter \chandra\ fluxes beyond $\approx$4~Ms will yield
larger gains in \xray\ source densities than previous increases in depth.  X11
showed that 
exposures of the \cdfs\ region will yield substantial gains in sensitivity.
For example, a feasible $\approx$10~Ms exposure would permit a factor of
$\approx$1.8 increase in depth (in terms of flux limit) over the current
$\approx$4~Ms exposure.  Therefore, exploring the normal-galaxy dominated
regime of the number-counts distribution is a sensible goal that may be pursued
in future surveys.  At present, the only \xray\ mission capable of reaching
these fluxes is \chandra\ (see below), and this will likely be the case for at
least the next two decades.

In Figure~16, we provide an expanded view of the SB number counts, and their
contributions by extragalactic source type (AGNs and normal galaxies), in the
faint flux regime.  Our best-fit number-counts parameterizations (based on
equation~5) have been indicated.  Direct extrapolation of these
parameterizations indicates that the number counts of normal galaxies will
overtake those of AGNs just below our current flux limit at
$\approx$\hbox{(3--5)}~$\times 10^{-18}$~\flux.  With a $\approx$10~Ms exposure
in the \cdfs, we could study number counts down to a sensitivity limit of
$\approx$$2.8 \times 10^{-18}$~\flux\ (see {\it vertical dotted line} in
Fig.~16), although we expect the faintest sources will have most-probable
fluxes of $\approx$$5.2 \times 10^{-18}$~\flux\ (see $\S$2.1 for distinction).
At such a depth, we would expect a factor of $\approx$1.7 increase in total
\xray\ source densities to $\approx$46,000~deg$^{-2}$ (based on extrapolation
of our best-fit number-counts parameterization), with normal galaxies making up
$\approx$57\% of the limiting source density.
After considering variations in sensitivity across the field, we expect that a
10~Ms exposure would yield \hbox{$\approx$1,020--1,080} total \xray\ detected
sources.

While \chandra\ is the only operating observatory capable of observing to the
ultra-deep regime studied here, the realization of future \xray\ mission
concepts like \genx\ and \smartx,\footnote{See
http://www.cfa.harvard.edu/hea/genx/ and http://hea-www.cfa.harvard.edu/SMARTX/
for more information about \genx\ and \smartx, respectively.} which have much
larger light-collecting areas and comparable or improved imaging resolution
over \chandra, would allow for the exploration of new populations of
extragalactic \xray\ sources, including new populations of distant ($z \simgt
0.6$) normal galaxies (see, e.g., Fig.~13).  Under its current specification, a
$\approx$4~Ms \smartx\ survey would detect SB sources with fluxes as low as
$\approx$$3 \times 10^{-19}$~\flux.  Direct extrapolation of our number counts
to these levels predict source densities of $\approx$500,000~deg$^{-2}$, with
$\approx$90\% contribution from normal galaxy populations.

\section{Summary}

In this paper, we have provided new quantitative measurements of the \xray\
number counts in the $\approx$4~Ms \cdfs\ using four bandpasses: 0.5--2~keV,
2--8~keV, 4--8~keV, and 0.5--8~keV (SB, HB, UHB, and FB, respectively).  Our
analyses focus on the faintest flux regimes only accessible by such ultradeep
\chandra\ surveys.  We draw the following key conclusions:

\begin{itemize}

\item We make use of a Bayesian approach, flux probability distributions, and
maximum-likelihood techniques to measure number counts for sources detected in
the $\approx$4~Ms \cdfs\ down to SB, HB, UHB, and FB flux limits of \sblim,
\hblim, \uhblim, and \fblim~\flux, respectively; these are factors of
\hbox{$\approx$1.9--4.3} times fainter than previous number counts
investigations.

\item At the flux limits of our survey, we reach source densities of \sbden,
\hbden, \uhbden, and \fbden~deg$^{-2}$ in the SB, HB, UHB, and FB,
respectively.  AGNs are the majority contributors to the total number counts at
all fluxes in all four bands (especially in the HB and UHB).  In the SB, we
reach AGN sky densities of $\approx$\agnden~deg$^{-2}$, the highest reliable AGN sky
density measured at any wavelength.

\item The normal-galaxy number counts rapidly rise, compared with AGNs, and at
the SB and FB flux limits, normal galaxies make up $\approx$46\% and
$\approx$43\% of the number counts, respectively.  At the limiting SB flux,
normal galaxies reach a sky density of $\approx$\galden~deg$^{-2}$, which is a
factor of $\approx$4.5 times higher than measured in the $\approx$2~Ms surveys.

\item Stated in terms of the intensity on the sky, \xray\ detected sources in
the $\approx$4~Ms \cdfs\ can account for $\approx$76--88\% of the CXRB.  We
estimate that the increased exposure from $\approx$1--2~Ms to $\approx$4~Ms has
resulted in an increase of $\approx$1--2\% resolved CXRB.

\item In the SB and HB, AGNs at high redshifts ($z \simgt 1.5$), with
moderate-to-significant obscuration ($N_{\rm H} \simgt 10^{22}$~cm$^{-2}$), and
relatively low intrinsic \hbox{0.5--8~keV} luminosity ($L_{\rm X} \simlt
10^{43}$~\lum) make increasingly important contributions to the AGN number
counts going to fainter fluxes, and at the flux limits, these sources make up
$\simgt$40\% of the overall AGN number counts.  These trends are broadly
consistent with those expected from the phenomenological models of G07.

\item In the SB and HB, late-type star-forming galaxies with $z \simlt 0.6$
provide majority contributions to the normal-galaxy number counts, with the
exception of the SB counts below $\approx$$10^{-17}$~\flux, where $z
\simgt$~0.6 star-forming galaxies dominate.  These galaxies are expected to
produce \xray\ emission from high and low mass \xray\ binaries, hot gas, and
young stars.  By contrast, passive early-type galaxies, with \xray\ emission
produced by low mass \xray\ binaries and hot gas, make up only small fractions
($\approx$5--20\%) of the SB and HB number counts near the survey flux limits.
These trends are well described by models that assume (1) the \xray\ power
output for late-type galaxies scales with SFR out to $z \approx 2$ with no
evolution in the scaling relation; and (2) the \xray\ power output in
early-type galaxies is due to a non-evolving hot gas emitting component and an
evolving (by $[1+z]^{2}$) low mass \xray\ binary component that both scale with
stellar mass.  Therefore, the rapidly rising normal-galaxy number counts can be
attributed primarily to the evolution of the physical properties of galaxies
(i.e., star formation rate and stellar mass) and not \xray\ scaling relations.

\item Extrapolation of our number counts to lower fluxes suggests that an
\xray\ observation reaching flux levels just below those probed by the
$\approx$4~Ms \cdfs\ (at SB fluxes of $\approx$[3--5]~$\times 10^{-18}$~\flux)
would result in an \xray\ sky with normal galaxies dominating the source
density at the sensitivity limit.  Since the normal-galaxy number counts are
now comparable with AGNs at the $\approx$4~Ms flux limit, and are expected to
continue to rise rapidly at fainter fluxes, more sensitive surveys will have
relatively large yields in increased source densities.  We show that a 10~Ms
\cdfs\ would yield \hbox{$\approx$1,020--1,080} total sources and would result in a
limiting SB source density of $\approx$46,000~deg$^{-2}$ with normal galaxies
dominating the number counts (providing $\approx$57\% of the total number
counts) at the flux limit.

\end{itemize}

\acknowledgements

We thank the anonymous referee for reviewing the manuscript and providing
useful suggestions.  We thank Andy Fabian for useful discussions and
acknowledge James Aird, Hermann Brunner, Fabrizio Fiore, Simonetta Puccetti,
and Shaji Vattakunnel for sharing data; these contributions have helped the
quality of this paper.  We gratefully acknowledge financial support from the
Einstein Fellowship Program (B.D.L.), CXC grant SP1-12007A and NASA ADP grant
NNX10AC99G (Y.Q.X. and W.N.B.), the Science and Technology Facilities Council
(D.M.A.), Financiamento Basal, CONICYT-Chile FONDECYT 1101024 and FONDAP-CATA
15010003, and CXC grant SAO SP1-12007B (F.E.B.), and ASI-INAF grants I/088/06
and I/009/10/0 (A.C., R.G., C.V.).  

\appendix

\section{Recovery Fraction Corrections}

As noted in $\S$2.2, the X11 \cdfs\ source catalog was generated following a
two-step approach, which entailed (1) running {\ttfamily wavdetect} to form an
initial list of candidate sources and (2) assessing the probability of
detection using {\ttfamily AE}.  Our methods described in $\S$~2 for computing
number counts generally account for biases due to completeness and the
Eddington bias provided our catalogs are complete to a specific {\ttfamily AE}
source selection probability.  Therefore, we need to account for
incompletenesses in the number counts that result as a consequence of our
two-step cataloging approach.  In this section, we summarize our approach for
computing this correction, which has been implemented in equation~5 as the
quantity $C^\prime$.

We began by generating 200 mock \cdfs\ images in the four bandpasses (i.e., 800
images), which each contained 700 \xray\ sources artificially implanted.  These
images were constructed as follows.  Each source was assigned a random right
ascension and declination within the area of the $\approx$4~Ms \cdfs\ as
defined by the exposure maps (see $\S$~3 of X11) and was given a random number
of counts between $\approx$1--100~counts.  The counts from each source were
added to a blank image canvas using the {\ttfamily marx}\footnote{See
http://space.mit.edu/cxc/marx/ for {\ttfamily marx} simulator details}
(version~4.5) ray-tracing code.  In this procedure, the PSF of each source was
modeled assuming a roll angle and aim point that were selected from one of the
54 observations that make up the cumulative $\approx$4~Ms \cdfs\ exposure (see
Table~1 of X11).  Selection of the roll angle and aim point for each source was
done probabilistically with the probability of selection being directly
proportional to the exposure time of each of the 54 \cdfs\ observations.  Mock
images for each of the four bandpasses were then created by adding background
counts from the respective cumulative background images (see $\S$7.1 of X11) to
the image canvases.

For each of the 700 sources in each of the 800 images we performed circular
aperture photometry using a circular aperture with a radius encompassing
$\approx$90\% of the encircled-energy fraction.  The number of counts measured
for each source was then compared with the number of counts needed for a source
detection (see discussion of sensitivity maps in $\S$2.1) to see if it would
satisfy the criterion adopted in equation~1 for source detection.  We then
searched each of the 800 images using {\ttfamily wavdetect} at a false-positive
probability threshold of $10^{-5}$ and made {\ttfamily wavdetect} source
catalogs for each image.  For each bandpass, we combined all 200 catalogs of
700 sources and computed the count rate $\phi$ and off-axis angle $\theta$
dependent fraction of sources that satisfied our binomial probability selection
criterion (i.e., equation~1) that were also detected by {\ttfamily wavdetect}
(hereafter, recovery fraction $f_{\rm recov} \equiv 1/C^\prime$).  

In Figure~A1, we show the $f_{\rm recov}$ versus $\phi$ for off-axis angle
intervals going to $\theta \approx 9$~arcmin, where the image is roughly
radially contiguous.  We found that the data for $f_{\rm recov}$ could
be represented successfully using the analytic form $f_{\rm recov}(\phi,\theta)
= 1/(1 + \exp[-\delta(\theta) \{\phi - \xi(\theta)\}])$, where $\delta(\theta)$
and $\xi(\theta)$ are fitting constants that vary with off-axis angle and
bandpass.  In $\S$~2.2, we made use of this relation in equation~6, where we
adopt $C^\prime = 1/f_{\rm recov}$.  We note that this correction has the
advantage in that it is independent of the Bayesian priors discussed in $\S$~2
and does not depend on the shape of the number counts distribution near and
below the flux limits.

%
%
\begin{figure*}
\figurenum{A1}
\centerline{
\includegraphics[width=16cm]{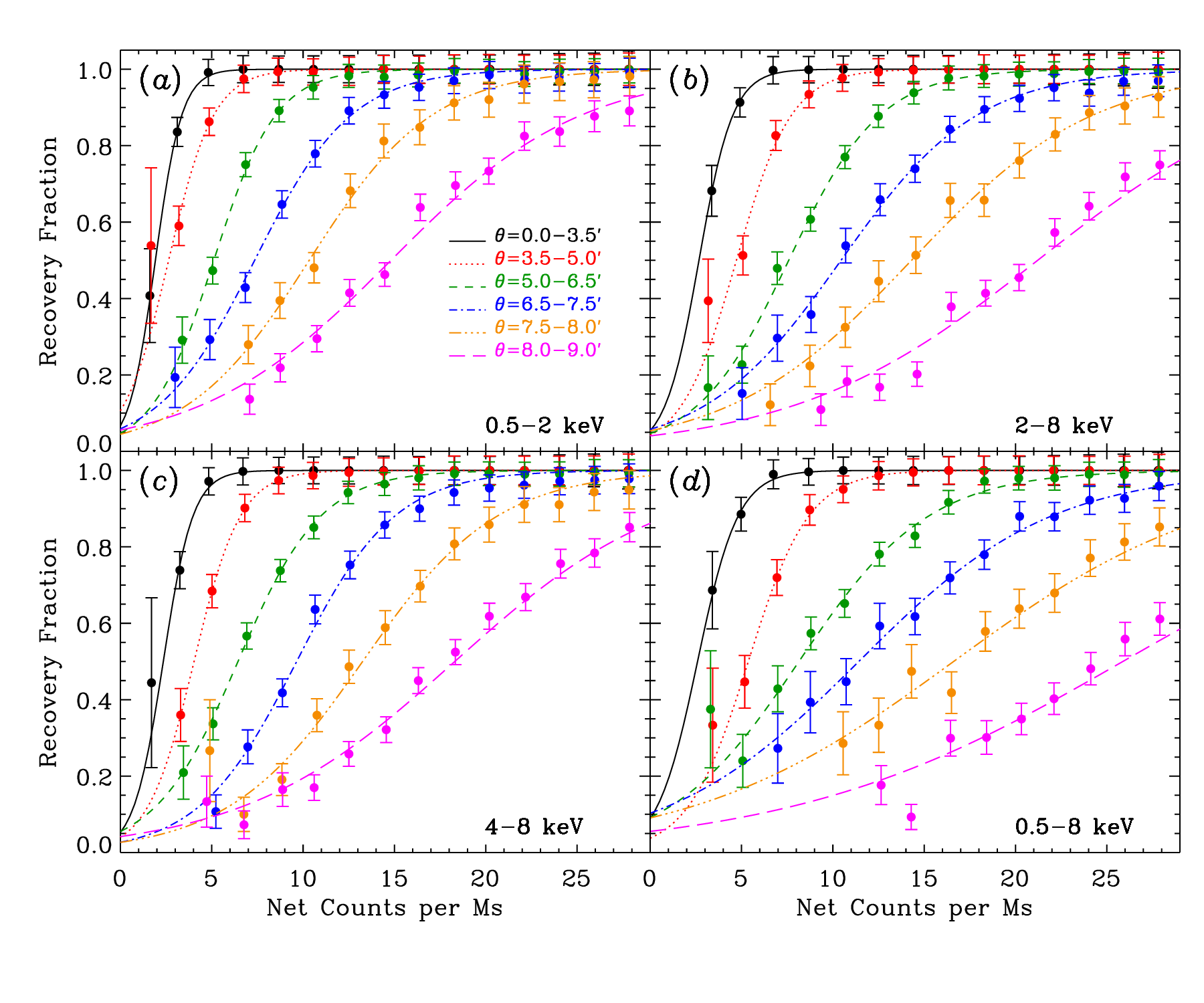}
}
\caption{
Fraction of simulated sources satisfying the binomial probability selection
criterion of equation~1 that were also detected using {\ttfamily wavdetect} at
a false-positive probablity threshold of $10^{-5}$ (i.e., $f_{\rm recov}$; {\it
filled circles with error bars\/}) for the four bandpasses.  Each color
represents a different off-axis angle interval as annotated on the plot.  Our
best-fit models, as described in the Appendix, have been plotted as curves (see
annotation); these models have been used to correct for incompleteness in our
number count computations in $\S$~2.
}
\end{figure*}

%

%

\end{document}